\documentclass[]{agujournal2018}
\usepackage{apacite}
\usepackage{url} 

\draftfalse

\journalname{JHM}

\usepackage{graphicx}
\usepackage{natbib}
\usepackage{MnSymbol}

\usepackage{textcomp}
\usepackage{amsthm}
\usepackage{mathptmx}

\usepackage{array}
\usepackage{multirow}
\usepackage{siunitx}
\usepackage{booktabs}
\usepackage{placeins}
\DeclareSymbolFont{cmsymbols}{OMS}{cmsy}{m}{n}
\SetSymbolFont{cmsymbols}{bold}{OMS}{cmsy}{b}{n}
\DeclareSymbolFontAlphabet{\mathcal}{cmsymbols}

\usepackage{soul}

\begin{document}
\correspondingauthor{Zeinab Takbiri}{takbi001@umn.edu}

\title{A Nested K-Nearest Prognostic Approach for Microwave Precipitation Phase Detection over Snow Cover}

\authors{Zeinab Takbiri \thanks {Department of Civil, Environmental and Geo-Engineering and St. Anthony Falls Laboratory, University of Minnesota, Minneapolis, USA.}, Ardeshir Ebtehaj \thanks {Department of Civil, Environmental and Geo-Engineering and St. Anthony Falls Laboratory, University of Minnesota, Minneapolis, USA.}, Efi Foufoula-Georgiou \thanks{Department of Civil and Environmental Engineering, University of California, Irvine, CA, USA}, Pierre-Emmanuel Kirstetter \thanks{Advanced Radar Research Center, University of Oklahoma, and National Severe Storms Laboratory, Norman, Oklahoma}, and F. Joseph Turk \thanks{Jet Propulsion Laboratory, California Institute of Technology, Pasadena, California}}

\begin{abstract}
Monitoring changes of precipitation phase from space is important for understanding the mass balance of Earth's cryosphere in a changing climate. This paper examines a Bayesian nearest neighbor approach for prognostic detection of precipitation and its phase using passive microwave observations from the Global Precipitation Measurement (GPM) satellite. The method uses the weighted Euclidean distance metric to search through an a priori database populated with coincident GPM radiometer and radar observations as well as ancillary snow-cover data. The algorithm performance is evaluated using data from GPM official precipitation products, ground-based radars, and high-fidelity simulations from the Weather Research and Forecasting model. Using the presented approach, we demonstrate that the hit probability of terrestrial precipitation detection can reach to 0.80, while the probability of false alarm remains below 0.11. The algorithm demonstrates higher skill in detecting snowfall than rainfall, on average by 10 percent. In particular, the probability of precipitation detection and its solid phase increases by 11 and 8 percent, over dry snow cover, when compared to other surface types. The main reason is found to be related to the ability of the algorithm in capturing the signal of increased liquid water content in snowy clouds over radiometrically cold snow-covered surfaces.
\end{abstract}

\section{Introduction}
More than two billion people rely on glacier and snowmelt for their water supply \citep{Mankin2015}. Snowfall accounts for approximately 30 to 90 percent of the global precipitation over mid- to high-latitudes \citep{Levizzani2011} and is the main input to the accumulation processes of snowpack and glaciers \citep{Radic2014}. In recent decades, snowpack reservoirs have declined and are projected to further decline in the 21\textsuperscript{st} century \citep{Karl1993, Mote2005, Pederson2011}. Thus, global monitoring of snowfall from space is key for improved understanding and prediction of ongoing changes in the cryosphere and the implications for sustainable management of water and food resources\,---\,especially in mountainous areas of the world. 

In the past three decades, significant progress has been made in microwave precipitation retrieval as part of the Tropical Rainfall Measuring Mission (TRMM) satellite in 1997 \citep{Kummerow1998}. The launch of the Global Precipitation Measurement (GPM) core satellite \citep{Kidd2011, Hou2014} has provided a unique opportunity for improved understanding of mid-latitude precipitation and its phase change beyond what the TRMM satellite could offer \citep{Skofronick-Jackson2017}.

The snowfall microwave scattering signal can be captured at frequencies above 80 GHz as these high frequencies are more sensitive to ice scattering compared to lower frequencies, which largely respond to variations of surface emissivity \citep{Kulie2010,Skofronick-Jackson2011, Gong2017,You2017}. Among high-frequency channels, \citet{Bennartz2003} found that frequencies around and above 150 GHz provide a strong polarization signal for snowfall detection \citep{Gong2017, You2017, Panegrossi2017}. 

Remote sensing of snowfall is among the most challenging tasks in precipitation retrieval algorithms \citep{Bennartz2003,Skofronick-Jackson2004,Noh2009,Kongoli2015}. Detection of snowfall is challenging because it involves complex and dynamic interactions between the snowfall scattering signal and the surface. First, compared to rainfall, the snowfall backscattering is much weaker \citep{Grody1991,Kim2008,Kulie2010} and depends on more complex microphysical characteristics snowfall such as shape and density of snowflakes \citep{Liu2008a,Petty2010,Skofronick-Jackson2011}. These characteristics are difficult to accurately parameterize as of today. Second the already weak snowfall scattering signal tends to be masked by the increased atmospheric emissivity and liquid water content in precipitating conditions \citep{Liu2013, Wang2013,Panegrossi2017}. Third, changes in surface emissivity due to snow accumulation on the ground can significantly alter the snowfall microwave signal. Dry snow cover scatters the upwelling surface radiation at frequencies above 20 GHz \citep{Ulaby1980,Hallikainen1987} similar to the snowfall \citep{Grody2008}. As a result, the snowfall microwave signature gradually weakens as snow accumulates on the ground \citep{Ebtehaj2017}. The snow-cover scattering evolves with time as a function of snow-cover metamorphism. For example, a small amount of liquid water content (e.g., 2\%) significantly reduces the snow-cover scattering and increases its absorptivity \citep{Stiles1980, Hallikainen1986, Hallikainen1987}. Hence, snow cover has a time-varying effect on snowfall upwelling signal.

Physical and empirical approaches have been developed for microwave retrievals of snowfall. \citet{Skofronick-Jackson2004} presented a physical method to retrieve snowfall during a blizzard over the eastern United States using high-frequency observations from the Advanced Microwave Sounding Unit (AMSU-B) instrument. \citet{Kim2008} simulated atmospheric profiles of a blizzard storm with the mesoscale MM5 model and a delta-Eddington type radiative transfer (RT) model to produce a storm-scale database for snowfall retrieval using AMSU-B observations. \citet{Noh2009} used a large number of snowfall profiles from airborne, surface and satellite radars, as well as an atmospheric RT model \citep{Liu98} to generate a regional database for snowfall retrievals using the AMSU-B data. The study used the NESDIS Microwave Land Surface Emissivity Model \citep{Weng2001} to provide surface emissivity as an input to the RT model. The largest retrieval error were found to be over snow-covered surfaces.

Empirical passive microwave snowfall retrieval algorithms largely rely on ancillary data of precipitation radar and air temperature profile. A family of these algorithms relies on thresholding the brightness temperature at different channels \citep[e.g.,][]{Staelin2000,Chen2003,Kongoli2003}. For example, \citet{Kongoli2015} developed a statistical approach that partitions high-frequency brightness temperatures ($\geq{89}$ GHz) into two distinct warm and cold weather regimes by thresholding the brightness temperature at 53 GHz.

Another class of empirical approaches relies on Bayesian techniques. These techniques use a database or a look-up table that relates brightness temperatures of snowing clouds to the radar snowfall observations along with the atmospheric temperature profile. As an example, \citet{Liu2013} used matched observations from the CloudSat Profiling Radar (CPR), the AMSU-B, and the NOAA'S Microwave Humidity Sounder (MHS). More recently, \citet{Sims2015} used the CloudSat radar and multiple ground-based reanalysis data, including near-surface air temperature, atmospheric moisture, low-level vertical temperature lapse rate, surface skin temperature, surface pressure, and land cover types to diagnose precipitation phase partitioning. This algorithm is deployed in the GPM operational precipitation retrievals \citep{Kummerow2015}. It is worth noting that most of these algorithms use reanalysis wet-bulb temperature that exhibits the strongest correlation with the precipitation phase \citep{Matsuo1981, motoyama1990simulation, Lundquist2008, Kienzle2008, Ye2013}. However, the reanalysis data are often available at coarse spatial scales with significant uncertainty, which hamper the applicability for accurate detection of snowfall \citep{Harpold2017}.

In this paper, we examine a prognostic Bayesian \textit{k}-nearest neighbor (KNN) algorithm that strictly relies on observed passive microwave brightness temperatures and does not use any online reanalysis data of temperature and moisture profiles. This approach is based on a weighted distance metric applied on an a priori database to detect overland precipitation phase. The a priori database is populated with combined radar-radiometer observations from the GPM satellite. This database is then stratified using data from the Moderate Resolution Imaging Spectroradiometer (MODIS) sensor to account for effects of the background snow-cover emission. We demonstrate that the algorithm shows improved skill in detection of snowfall over snow cover and can predict the likelihood of precipitation phase changes in the atmospheric boundary layer, which is not well observed by the GPM radar. 

In summary, the presented algorithm isolates a few physically relevant candidate vectors of brightness temperatures in the database via a weighted Euclidean distance. Using these isolated candidates, the method detects the precipitation and its phase, based on a probabilistic decision rule. To test the performance of the proposed approach, the database is populated by merging the outputs of both passive \citep{Sims2015} and active \citep{Iguchi2010} GPM products using all overland observations from June 2015 to May 2016. We compare the results with the ground-based Multi-Radar Multi-Sensor (MRMS) data over the Conterminous United States (CONUS) \citep{Zhang2011, Zhang2016}. The outputs of a high-fidelity mesoscale simulation model are also used for further evaluation of the results over high altitudes, during the Olympic Mountains Experiment in 2015 \citep[OLYMPEx,][]{Houze2017}. 

The paper is structured as follows: Section 2 briefly describes the database and the phase detection method used on the operational GPM radar and radiometer products. Section 3 elaborates on the effects of snow cover on passive microwave signal of snowfall at different frequency channels by analyzing a large dataset of GPM observations. Section 4 explains the proposed KNN algorithm followed by the results presented in Section 5. Concluding remarks and future directions of the research are discussed in Section 6. 

\section{Database Description}
The dual-frequency precipitation radar (DPR) aboard the GPM core satellite detects precipitation reflectivity at Ka- (35 GHz) and Ku-band (13.6 GHz). The GPM Microwave Imager (GMI) captures the upwelling emission/scattering signals of the surface and the atmosphere at 13 frequency channels ranging from 10 to 183 GHz. On the one hand, observations by the DPR and the GMI high-frequency channels ($>80$~GHz) provide information about the microwave signature of precipitation and more specifically about snowfall ice scattering. On the other hand, observations by the low-frequency channels ($>80$~GHz) add information about the land surface characteristics that leads to improved detection skill by the presented algorithm. This study uses level-II near-surface precipitation phase retrieval from DPR (active) product (2A-DPR-V04, Normal Scan), GMI (passive) product (2A-GPROF-V04) and the level 1B calibrated GMI brightness temperatures. 

In DPR level-II, the precipitation phase is determined by the dual-frequency retrieval approach that uses the differential attenuation between the Ku- and the Ka-band reflectivity values \citep{Iguchi2010,Iguchi+Seto2012}. The differential attenuation method ingests ancillary atmospheric profile data such as air temperature, pressure, and the microphysical parameterization of the snow and rain particle size distribution. The DPR surface retrieval is inferred from the near-surface reflectivity observations in the clutter-free region. Above relatively flat surfaces, the altitude of this region varies from 1 to 2 km from nadir to the edge of the DPR swath. The depth of this region is often increased over complex terrains. As a result, any precipitation within the ground clutter region cannot be detected by the radar. Moreover, DPR has limited capability to detect light precipitation with a rate below $\sim$\,\SI{0.2}{mm.h^{-1}} \citep{Hou2014, Kubota2014}.

Unlike the DPR that provides range-resolved information about the precipitation backscattering, the GMI observes an integration of precipitation scattering in a continuum that extends from the land surface to the top of the atmosphere. As previously explained, the current operational algorithm for passive detection of precipitation phase relies on thresholding of the near-surface wet-bulb temperature \citep{Sims2015}. The wet-bulb temperature is processed offline from reanalysis of ancillary data, which often suffer from different sources of uncertainty, especially due to its coarse resolution over topographic features and structurally complex land surfaces \citep{Li2008}.

For implementing and testing the proposed algorithm, we create a reference product (REF) for precipitation occurrence and phase change. This REF product is based solely on the occurrence information from the DPR data. For determining the precipitation phase, we use the inner-swath phase information from both GPM active and passive products. None of these products provides direct phase estimation; however, each has unique information based on the atmospheric and surface conditions. Specifically, the REF product determines the phase by applying a logical operator to both active and passive products. The radar phase retrieval is reported as solid, liquid, and mixed, while the phase probability in GPROF is from zero (solid) to one (liquid). We therefore first, discretize the GPROF phase probability into solid (phase probability less than 0.5) and liquid (phase probability greater than 0.5) to match the radar phase. Second, we assign the phase of REF precipitation as solid or liquid when both active and passive phases are solid or liquid. Otherwise, the phase is labeled as mixed. Therefore, the mixed phase in the REF product refers to those cases where the precipitation phases from the passive and active products do not agree and thus should not be literally interpreted. By combining the active and passive phase information through this logical rule, we implicitly address the limitations of DPR in identifying precipitation phase change in the ground clutter region which overlaps with the boundary layer.

It is important to note that the so-called mixed category depends on the threshold (0.5), used for discretization of the passive phase. Understanding the impacts of this threshold on the retrieval requires a thorough investigation outside the scope of this study. It is worth noting that choosing this threshold results in 12\% of mixed phase data in the REF product, in which 10\% corresponds to liquid passive phase and solid active phase (scenario 1) and 2\% otherwise (scenario 2). The first scenario might be related to those conditions where the melting layer is in the clutter region. The second scenario may be related to a temperature inversion near the surface that causes a refreezing of precipitation.

The MODIS daily snow-cover fraction \citep[MOD10A1,][]{Hall2002} and land surface skin temperature \citep[MOD11A1,][]{Wan2014} are added to the database. The data are used from the MODIS sensor on board the Terra satellite. The MODIS snow cover and surface temperature data are available at a resolution of 500 and 1000 meters, respectively. We assume that the total daytime snow-cover fraction does not change significantly between consecutive overpasses of the GPM and Terra satellites within one day. Note that this assumption could give rise to some degree of uncertainty when the data are matched with nighttime precipitation. We consider a 5\,km DPR pixel as a snow-covered surface when more than 50\% of the enclosed high-resolution snow fraction data indicates the presence of snow cover on the ground. It is also assumed that the temperature does not vary significantly within a 5\,km DPR pixel and is considered to be the average of the cloud-free MODIS temperature data. As the liquid water content of global snowpack is not available, we define dry (wet) snow when the skin and air temperature are both below (above) \SI{0}{\celsius} \citep{Baggi2009}.

To account for atmospheric radiometric signals, we also added the integrated liquid and ice water content of the clouds, as well as the integrated water vapor content of the atmospheric column from the second version of the Modern-Era Retrospective analysis for Research and Applications \citep[MERRA-2-M2I1NXASM,][]{Gelaro2017}. The MERRA-2 data are hourly single-level diagnostic products at \SI{0.625}{\degree}$\times$\SI{0.5}{\degree}, which are derived from the version 5 of the NASA Goddard Earth Observing System (GEOS-5) and the Atmospheric Data assimilation system (ADAS).

To form the database with a uniform spatial sampling density, the GMI brightness temperatures and the MERRA-2 reanalysis data are mapped onto the DPR grids and scanning time using the spatial nearest neighbor interpolation and temporal linear interpolation techniques. The high-resolution MODIS snow-cover data are mapped onto and then averaged over the nearest DPR grids. We collect and process two consecutive years of data, from June 2014 to May 2016, which lead to a database with more than \num{5e9} matched data pairs. The data from the first year (June 2014 to May 2015) are applied to build the database and the data from the second year (June 2015 to May 2016) are used to test the proposed algorithm.

\section{The effect of snow cover on terrestrial snowfall signal}
\begin{figure*}
 \centering
 {\includegraphics[width=\textwidth]{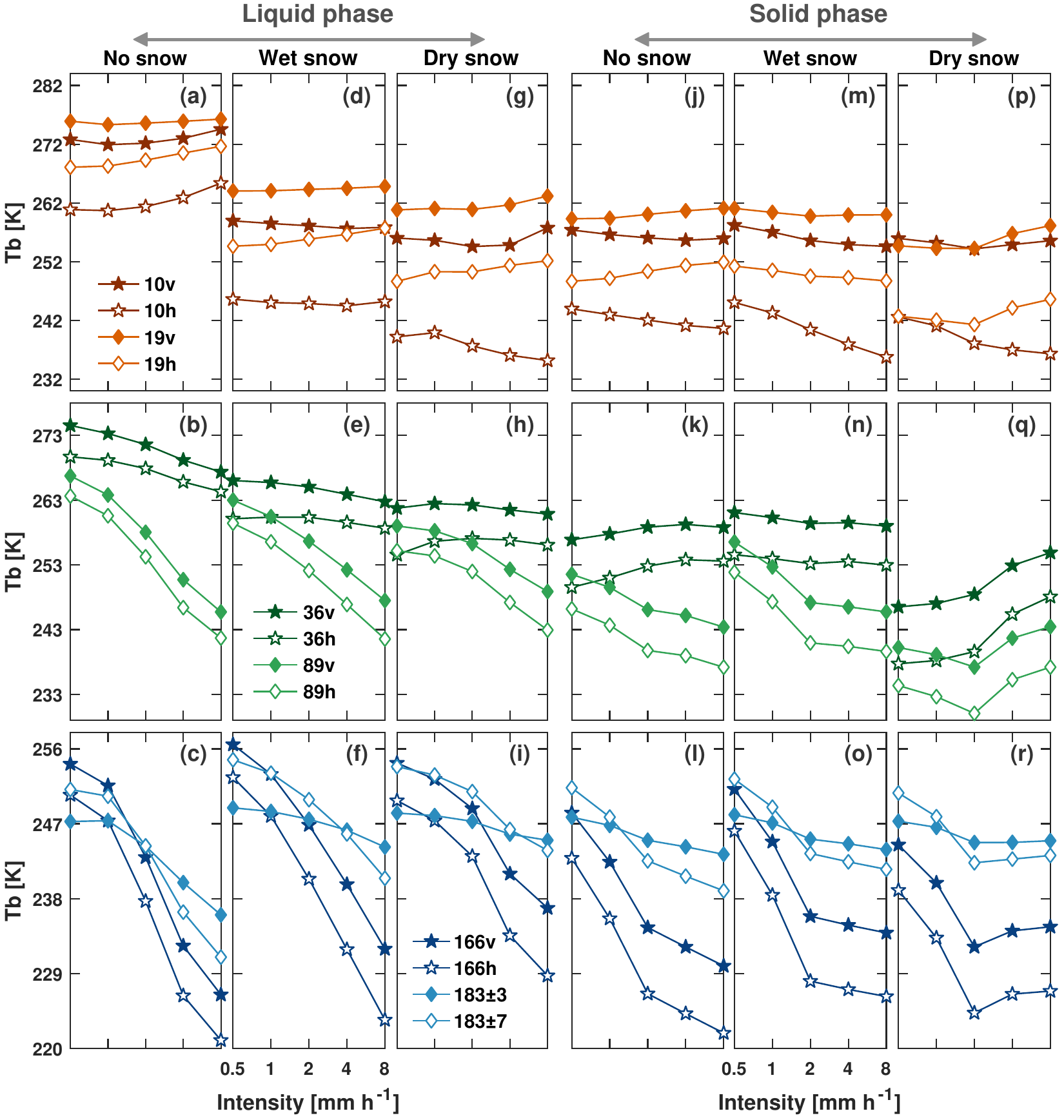}}
\caption{Variation of mean brightness temperatures (June 2014 to May 2016) in response to changes in precipitation intensity for different land-atmosphere classes including the liquid (a-i) and solid (j-r) precipitation phase over the ground (no snow), wet and dry snow cover. The bins are five logarithmically (base 2) spaced intervals with the width of \SI{1}{mm.h^{-1}} in the log-space.}\label{fig01} 
\end{figure*}

Precipitation spectral signatures and their dependence on snow-cover scattering are studied by analyzing the entire dataset (June 2014 to May 2016) for three surface types (ground without snow cover, wet snow, and dry snow) and for both liquid and solid phases of precipitation. Each land-atmospheric class is further divided into 5 bins of precipitation intensity $r$ with equal logarithmic width, \mbox{$\log_{2}(r_{i+1}/r_{i})=1$}, centered at \numlist{0.5;1;2;4;8}\,\si{mm.h^{-1}}. We first quantify the effects of snow cover on the precipitation signal over each surface type by calculating the mean values of the brightness temperatures for different precipitation phases and intensities at frequency bands 10--19, 36--89, 166, 183$\pm3$, and 183$\pm7$ GHz (Fig.~\ref{fig01}). Then, we demonstrate that the snowfall signal exhibits a weaker scattering signal than rainfall and reveal that there exists a non-unique relationship between the brightness temperatures and snowfall rate over snow-covered surfaces. Lastly, we highlight why precipitation phase detection could be less uncertain over dry than wet snow cover using the presented approach.

The first three columns in Fig.~\ref{fig01}\,a--i focus on the signatures of rainfall over land surfaces with no snow cover, wet snow cover, and dry snow cover, where both active and passive products indicate liquid phase. The signatures over the ground with no snow cover are mainly affected by the upwelling surface emission, the upwelling emission by cloud liquid water path, as well as scattering by the cloud ice particles and large raindrops. As it is well understood, due to strong background emission at frequencies 10--36 GHz, the overland precipitation microwave signal is difficult to be separated from the surface contributed signal in these channels. For example, due to the rainfall emission, the mean brightness temperature at 10h GHz only increases by less than 5\,K as the intensity increases from 0.5 to 8\,\si{mm.h^{-1}} (Fig.~\ref{fig01}\,a). 

On average, we observe that over all three land surface types, the brightness temperatures monotonically decrease for frequencies above 80 GHz as the rainfall intensity increases. However, the significance of scattering decreases over snow-covered surfaces (Fig.~\ref{fig01}\,d-i). For example, at 89 and 166 GHz, the average decrease of brightness temperature per 1\,\si{mm.h^{-1}} increase in rainfall intensity is about 3.0 and 3.6\,K (Fig.~\ref{fig01}\,d,\,g), while these rates are around 1.2 and 2.3\,K over the dry snow cover (Fig.~\ref{fig01}\,h,\,i). As expected, the $183\pm3$ GHz is the least sensitive channel to the changes of rainfall rate. This channel becomes almost insensitive to the rainfall intensity when the snow accumulates on the ground and exhibits less than 0.2\,K of cooling effect per unit rainfall (Fig.~\ref{fig01}\,i).

\begin{figure*}[h!]
 \centering
 {\includegraphics[width=\textwidth]{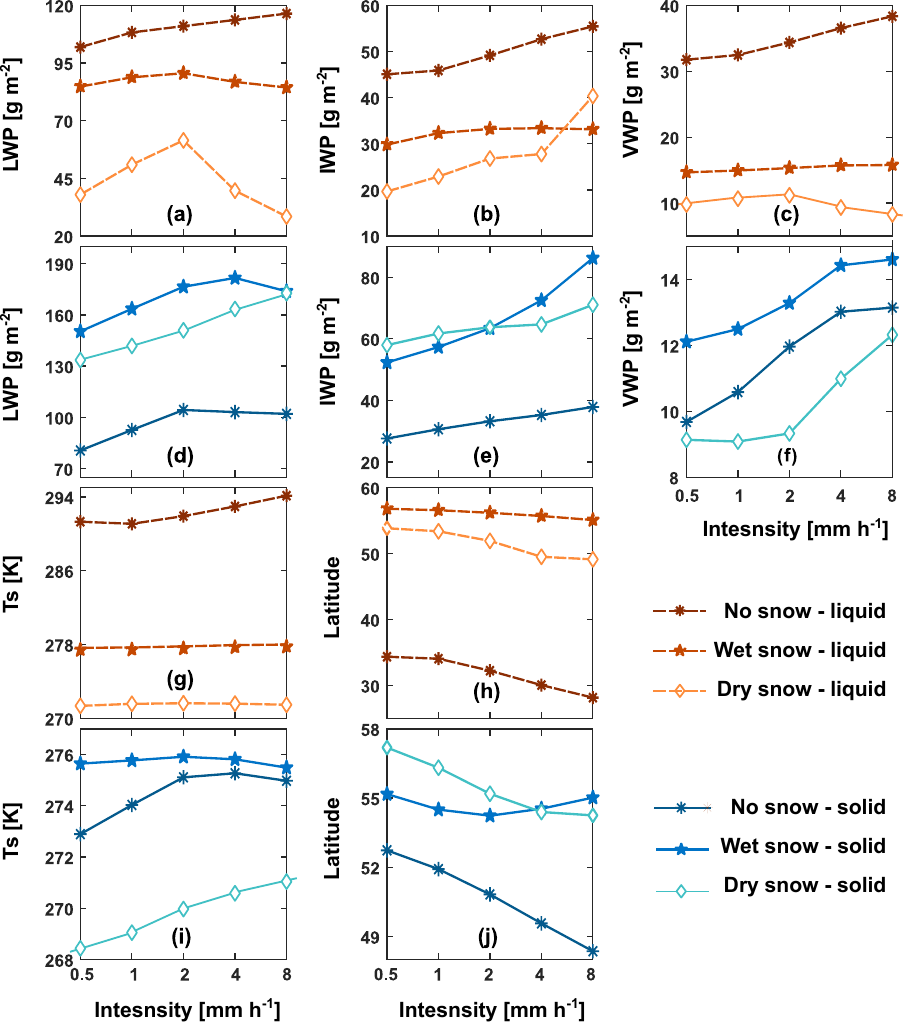}}
\caption{Average variations of the cloud liquid water path (LWP) (a,/,d); the cloud ice water path (IWP) (b,/,e); the water vapor path (WVP) (c,/,f); the skin temperature (Ts) (g,/,i); and latitudes (h,j) against the precipitation intensity. The ice and liquid water paths are extracted from the MERRA-2 data \citep[M2I1NXASM,][]{Gelaro2017} and the surface temperature data are from MODIS \citep[MOD11A1]{Wan2014} during June 2014 until May 2016. The intensity bins are the same as Fig.~\ref{fig01}.}\label{fig02} 
\end{figure*}

The last three columns in Fig.~\ref{fig01}\,j--r present brightness temperatures of snowfall over the three explained land surface types, where both active and passive products indicate solid phase. Similar to the overland rainfall, the emission and the scattering signals become more significant from low to high-frequency channels. Over the surfaces with no snow cover, when the snowfall intensity increases from 0.5 to 8\,\si{mm.h^{-1}}, the brightness temperatures at frequencies $\leq{36}$~GHz increase $\sim{6}$~\si{\kelvin} (Fig.~\ref{fig01}\,j). This warming could be due to increased cloud liquid water path (from 75 to 101\,\si{g.m^{-2}}, Fig.~\ref{fig02}\,a,\,d), water vapor path (from 9.5 to 13.1 {\si{kg.m^{-2}}}, Fig.~\ref{fig02}\,c,\,f), and surface temperature (from 273 to \mbox{\SI{274.2}{\kelvin}}, Fig.~\ref{fig02}\,g,\,i).

Due to the snowfall scattering, the average brightness temperature at 166 GHz frequency channel (Fig.~\ref{fig01}\,l,\,o,\,r) decreases about \numrange{14}{20}\,\si{\kelvin}, which corresponds to a cooling rate of \numrange{1.75}{2.50}\,\si{\kelvin} per unit snowfall rate. This observation reaffirms the importance of 166 GHz for snowfall retrieval compared to the 89 GHz channel \citep[see][]{Bennartz2003,Shi2010,Skofronick2013,You2017}. When the precipitation intensity increases from 0.5 to 8\,\si{mm.h^{-1}}, the average decrease in brightness temperatures at 166 (89) GHz is about \numrange{18}{30} (\numrange{10}{22})~\si{\kelvin} for rainfall and \numrange{10}{20} (\numrange{2}{9})~\si{\kelvin} for snowfall\,---\,over all examined land surface types. Therefore, the scattering signal weakens when the precipitation falls in the solid form; however, this weakening effect is less significant at 166 GHz than 89 GHz. In particular, over the ground with no snow cover, the signal becomes weaker approximately by 30\% and 57\% at 166 and 89 GHz, respectively, while these rates are 44 and 80\% over the dry snow cover. 

Observations demonstrate that the snowfall scattering signal decreases at frequencies $\geq89$ GHz when snow begins to accumulate on the ground. An interesting observation is the non-monotonic response of the observed brightness temperatures to the snowfall rate over snow-covered surfaces. For example, over the dry snow, the brightness temperatures at $\geq89$ GHz increase when the snowfall intensity varies from 2 to \SI{4}{mm.h^{-1}}, showing an irregular transition from a scattering to an emission regime (Fig.~\ref{fig01}\,q,\,r). Although less pronounced, a similar pattern is observed over the wet snow cover (Fig.~\ref{fig01}\,n,o).

The possible reasons for this anomaly could be related to an emission signal from either the atmosphere or the land surface. The atmospheric-related reasons can be due to the enhanced emission from the cloud liquid water and/or the water vapor path, both of them often increase with increasing snowfall intensity \citep{Liu2013,You2017,Ebtehaj2017}. The land surface-related causes largely correspond to the increased surface temperature and/or changes in the snow-cover physical properties. To find the most prominent contributing factor, we analyzed the variations of liquid, ice, and vapor water path derived from MERRA-2 data, the surface temperature derived from MODIS, and the mean snowfall intensity at different latitudes (Fig.~\ref{fig02}\,a-f). 

Over the ground with no snow cover, as the average precipitation intensity increases, the liquid and ice water path increase during rainfall and even more significantly during snowfall. Specifically, the liquid water path increases from 14 to 26\% (Fig.~\ref{fig02}\,a,\,d) and the ice water path increases about 23 and 37\% (Fig.~\ref{fig02}\,b,\,e) for raining and snowing events, respectively. Over dry snow cover, there is no evidence of any additional changes neither in liquid nor in ice water path that could cause the observed irregularity. Fig.~\ref{fig02}\,f shows that the water vapor path increases about 2.5\,\si{kg.m^{-2}} between snowfall intensities 2 and 8~\si{mm.h^{-1}} over the dry snow cover, which cannot be the main reason for the observed anomaly. The reason is that the sensitivity of the 166 GHz channel to variation of water vapor decreases significantly for snowfall intensities $>0.8$~\si{mm.h^{-1}} \citep{You2017}. Therefore, we speculate that the anomaly could be largely due to a surface effect.

The MODIS surface temperature data \citep{Wan2014} do not show any significant dependency on the rate of snowfall (Fig.~\ref{fig02}\,i). Therefore, we hypothesize that the detected emission could be due to either an unknown retrieval uncertainty or more likely, to the climatology of the snowfall and snow cover dynamics. The database shows that light but prolonged snowfall intensities ($<$~\SI{2}{mm.h^{-1}}) occur at latitudes above $>55$\si{\degree}~N over dry and thicker snow cover (Fig.~\ref{fig02}\,j). However, high intensity but less frequent snowfall is more likely to occur over lower latitudes with a thinner snow cover climatology. In other words, the high snowfall rates mostly represent the climatology of lower latitudes with thinner depth of snow cover, less volume scattering, and thus stronger surface emission than the thicker snow cover of higher latitudes.

\begin{figure*}[!ht]
 \centering
 {\includegraphics[width=\textwidth]{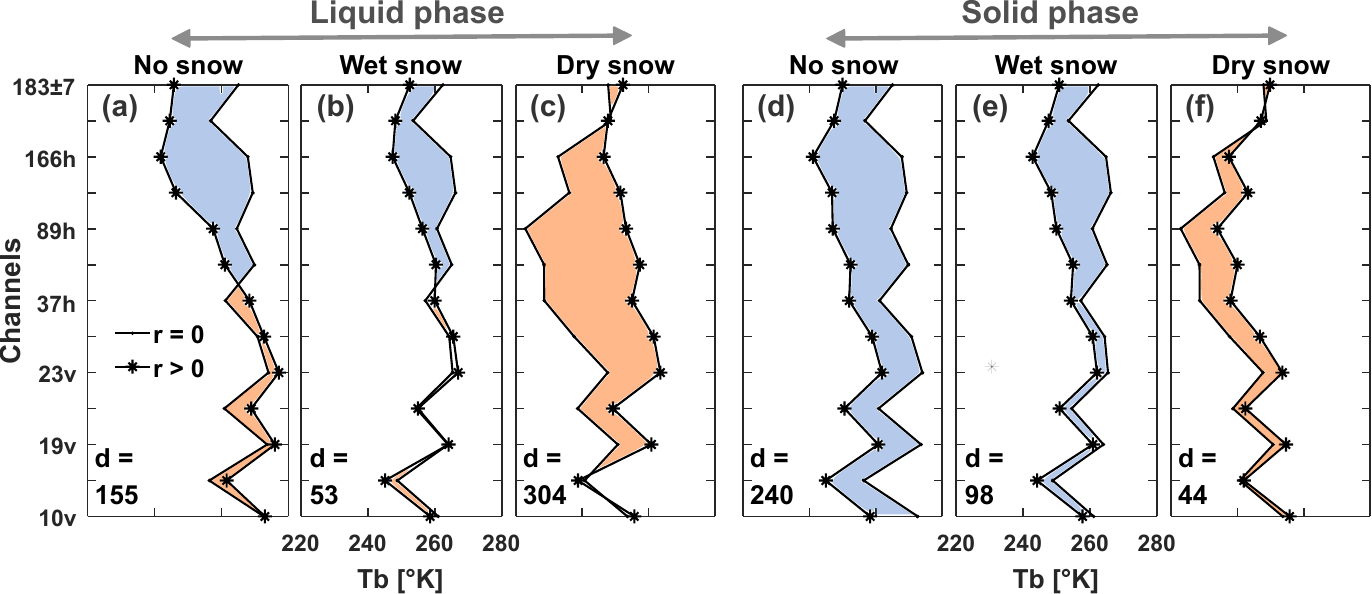}}
\caption{Average distance between vectors of mean brightness temperatures in the database from June 2014 to May 2016 for a clear-sky ($r=0$) and a near-surface precipitating atmosphere ($r>0$) with liquid (a-c) and solid phase (d-f) overland precipitation with no snow cover, wet snow cover, and dry snow cover. The blue and orange shaded areas indicate the cooling (scattering) and warming (emission) signals of precipitation. The mean root squared distance between the no precipitating (clear-sky) and precipitating atmosphere is also presented for each land-atmosphere class.}\label{fig03}
\end{figure*}

The above observations from Figs~\ref{fig01} and \ref{fig02} lead us to hypothesize that the distance between vectors of brightness temperature encodes a similarity metric that can be used to discriminate the precipitation from the background signal. A larger distance indicates larger radiometric dissimilarity that could be used for improved detection of the precipitation from the background signal. Using the database, we calculate the average distance between the vectors of brightness temperatures for the clear-sky (no precipitation) and precipitating atmosphere over the three land surface types (Fig.~\ref{fig03}). In this figure, the shaded areas in light blue (orange) represent the detected emission (scattering) signal. The key observation is that when the snow-cover scattering increases, the precipitation signal transitions from a scattering to an emission regime. The wet snow cover weakens the precipitation scattering as it is less emissive than the ground with no snow cover. However, the less emissive dry snow reveals the precipitation emission signal.

For the liquid phase, we can see that the rainfall scattering over the ground with no snow cover is manifested by a large distance between the high-frequency channels $\geq$~89~GHz, while the distance over lower frequency channels is insignificant (Fig.~\ref{fig03}\,a). This distance shrinks over the wet snow cover (Fig.~\ref{fig03}\,b), where the dominant precipitation signal is still due to its scattering over high-frequency channels. This shrinkage is largely explained because wet snow is not a strong scatterer and thus reduces slightly the surface emission and the high-frequency scattering of rainfall. However, when the surface emission is significantly reduced over the dry snow (Fig.~\ref{fig03}\,c), the emission of rainfall can be detected as a warming signal across almost all frequency channels. 
For the solid phase, the distance is relatively large between the background and precipitation signals when there is no snow on the ground (Fig.~\ref{fig03}\,d). This distance captures a shift across all frequency channels and a reduced polarization signal above 37 GHz. The shift is largely due to the differences between the surface temperature of clear-sky versus a snowing atmosphere, while the reduced polarization is chiefly due to diffused scattering of the snowflakes. Similar to the liquid precipitation, this distance shrinks when the ground is covered with wet snow, where the shift between the background temperature almost vanishes as the surface temperature increases. We can see that when the snowfall is occurring over dry snow, an emission signal is observed, chiefly in response to the increased liquid and water vapor paths \citep[see][]{Liu2013,You2015,You2016,Ebtehaj2017}. The MERRA-2 data indicate increases of $\sim$\,\SI{58}{g.m^{-2}} and \SI{4.8}{kg.m^{-2}} in liquid and vapor water paths, respectively, when snowfall occurs. This emission signal indirectly indicates the likelihood of precipitation by increasing the brightness temperatures rather than a direct physical signature of precipitation. Because of this emission signal, the vector of snowfall brightness temperatures becomes dissimilar to the surface emission, which could lead to improved snowfall retrievals over dry snow cover\,---\,if a proper distance metric is used to quantify the dissimilarity. 

\section{A Nested Nearest Neighbor Algorithm for Precipitation Phase Detection}

The nearest neighbor matching has been successfully utilized for passive microwave retrieval of rainfall using the TRMM data \citep{Ebtehaj2015,Ebtehaj2016} and for microwave mapping of flood inundation using the Special Sensor Microwave Imager/Sounder observations \citep{Takbiri2017}. In this section, we introduce a prognostic algorithm that relies on a nested {\it k}-nearest neighbor matching that finds the best representation of a query brightness temperature in the database to detect precipitation occurrence and phase. The criterion for matching relies on the hypothesis that similar vectors of brightness temperatures represent similar atmospheric profiles. In other words, an observed pixel-level vector of brightness temperature for a precipitating atmosphere is more similar to a collection of precipitating brightness temperatures in the database than those that refer to a non-precipitating atmosphere. Here, we define the similarity metric by a weighted Euclidean distance between the query vector of observed brightness temperatures and those stored in the a priori database, described in Section 2. 

\begin{figure*}[!ht]
 \centering
 {\includegraphics[width=0.85\textwidth]{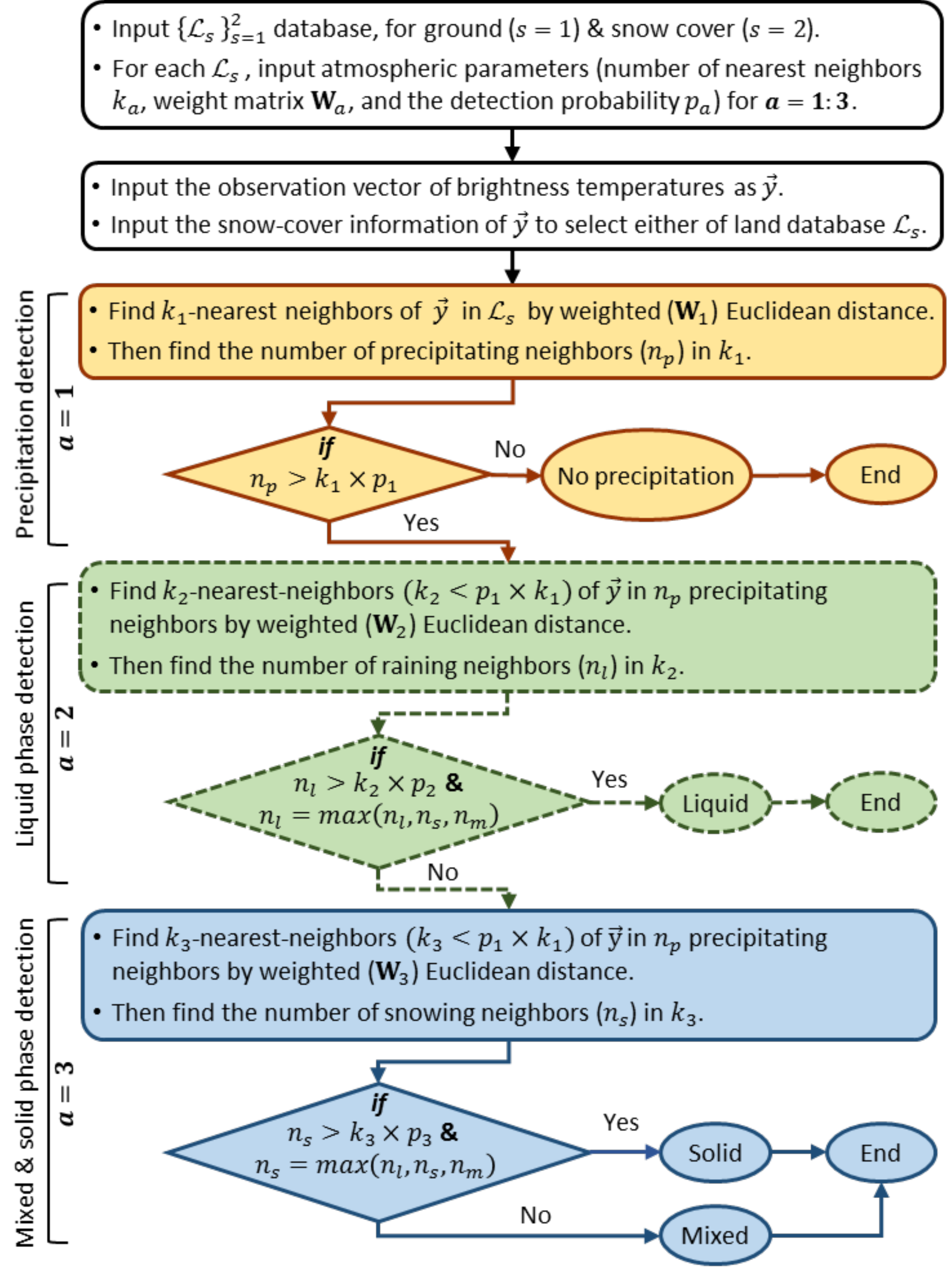}}\vspace{-4mm}
\caption{Algorithm flowchart of the proposed weighted \textit{k}-nearest neighbor (KNN) algorithm for detection of precipitation occurrence and phase.}\label{fig04} 
\end{figure*}

To set the notation, hereafter, the vector of brightness temperatures is denoted by $\vec{Tb}$ and the ancillary data containing information on the precipitation occurrence, phase, and snow cover, are represented by the vector ${\vec{r}}$. The database is pruned to contain balanced information over two different land-surface types $\left\lbrace{\mathcal{L}}\right\rbrace_{s=1}^{2}$ and four independent atmospheric conditions $\left\lbrace{\mathcal{A}}\right\rbrace_{a=1}^{4}$. The land surface types are defined only based on the presence ($s=1$) and absence ($s=2$) of snow cover, while the atmospheric conditions refer to the clear sky ($a=1$), liquid ($a=2$), solid ($a=3$), and mixed ($a=4$) precipitating atmosphere. 

To reduce the algorithmic complexity, we do not differentiate between the dry and wet snow cover in the database. Each land class consists of pairs of $\left\lbrace\left({\vec{Tb}}_m,\, {\vec{r}}_m\right)\right\rbrace_{m=1}^{M}$, where ${M}=2\times10^{7}$ are evenly distributed between clear and precipitating sky. Those pairs in the precipitating sky are also evenly distributed between raining $(\mathcal{A}_{2})$, mixed $(\mathcal{A}_{3})$, and snowing $(\mathcal{A}_{4})$ atmosphere. It is important to note that the pairs are randomly selected from their parent class to avoid any bias toward a specific land or atmospheric class.

In summary, for a given land surface type and a query vector of brightness temperatures ${\vec{y}}$, the algorithm relies on a 3-tuple $\left\lbrace\left(k_a,\mathbf{W}_a, p_a\right)\right\rbrace_{a=1}^{3}$, where ${k}_a$ is the number of nearest neighbors, $\mathbf{W}_a$ is the atmospheric weight matrix over each land surface type used in the weighted Euclidean distance $ d_{m}=\left(\vec{y}-\vec{Tb}_{m}\right)^{{\rm T}}\mathbf{W}_{a}\left(\vec{y}-\vec{Tb}_{m}\right)$, and $p_a$ denotes a detection probability measure. The weight matrix accounts for the relative importance of the channel combinations for detection of precipitation and its phases \citep{Ebtehaj2017}. 
Specifically, given the land surface types $\mathcal{L}_s$, after finding the $k$-nearest neighbors $\left\lbrace\left({\vec{Tb}}_i,{\vec{r}}_i\right)\right\rbrace_{i=1}^{k}$ of each query vector ${\vec{y}}$, a nested decision-making process is made to detect precipitation and its phase based on the probability measure $p_a$.

In the first step, the algorithm uses $\left(k_1, \mathbf{W}_1,\, p_1\right)$ to search for the $k_{1}$-nearest neighbors of $\left\lbrace{\vec{Tb}}_i\right\rbrace_{i=1}^{k_1}$ and the corresponding ancillary information in the database. Then, a binary decision is made to label the vector ${\vec{y}}$ as a precipitating observation, when the number of precipitating neighbors $n_p$ is greater than $p_1\times k_1$. For precipitating ${\vec{y}}$, the algorithm identifies the precipitation phase by running a new $k$-nearest neighbor search using $\left(k_{2}, \mathbf{W}_2, p_2\right)$ through those precipitating neighbors $\left\lbrace\left({\vec{Tb}}_j, {\vec{r}}_j\right)\right\rbrace_{j=1}^{n_p}$ that are found in the first step, where \mbox{$k_2<p_1\times k_1$}. Then, as explained before, a binary decision is made to label $\vec{y}$ as liquid precipitation, if the number of raining neighbors $n_l=\max(n_l,\,n_s,\,n_m)$ is greater than $p_2\times k_2$, where $n_m$ and $n_s$ are the number of mixed and solid precipitation elements among the $k_2$-nearest brightness temperatures $\left\lbrace{\vec{Tb}}_i\right\rbrace_{i=1}^{k_2}$. If those conditions are not satisfied, the algorithm continues similarly to find if the phase of $\vec{y}$ is solid or mixed. An algorithmic flowchart is presented in Fig.~\ref{fig04}.

\begin{figure*}[!ht]
 \centering
 {\noindent\includegraphics[width=0.9\textwidth]{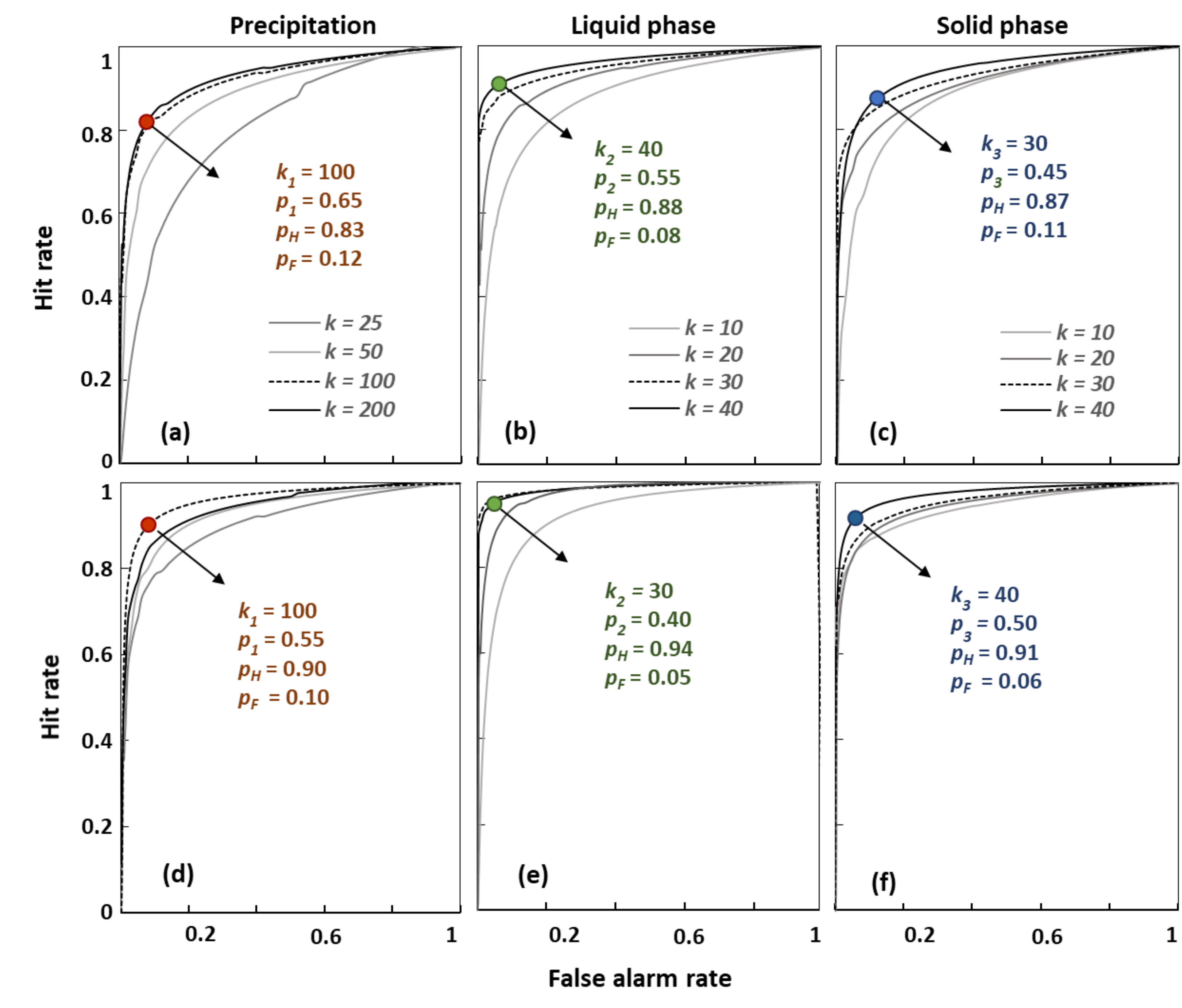}} 
\caption{Trade-off curves between the probability of hit ($p_H$) and false alarm ($p_F$) calculated with different numbers of \textit{k}-nearest neighbors for detection of the precipitation occurrence and phase over no-snow covered surfaces (top row: a--c) and snow-covered surfaces (bottom row: d--f). The optimal values of \textit{k} and the detection probabilities \textit{p} are given in each subplot.}\label{fig05} 
\end{figure*}

To determine the optimal values of the input parameters $k_a$ and $p_a$, we compute the receiver operating characteristic curves (ROC, Fig.~\ref{fig05}), which characterize the tradeoff between the hit and false alarm rates. The probability of hit is defined as the fraction of occurred events that were correctly detected, while the false alarm rate is a fraction of events that did not occur but were incorrectly detected by the algorithm. Let $a$ represent the number of correctly detected events, $c$ the number of missed events, $b$ the number of false detection, and $d$ the number of correct rejection. Then, the probabilities of hit and false alarm are defined as $\dfrac{a}{a+c}$ and $\dfrac{b}{b+d}$, respectively. The optimal value of $k_a$ is chosen based on the maximum area under the ROC curves \citep{Hanley1982}, while the best detection probability $p_a$ is chosen where the curvature of the ROC is maximum.

\section{Results and Validation}

To test the performance of the proposed approach, the terrestrial precipitation and its phase are retrieved over the inner-swath of the GMI overpasses from June 2015 to May 2016. As the phase outputs of the algorithm are discrete values for solid (0), mixed (0.5), and liquid (1), the temporal mean values associated with these phases could reveal the overall sensitivity of the algorithm to the seasonal variations of surface temperature and emissivity. To that end, the phase indices are averaged at orbital levels over the summer and winter for the nested {\it{k}}-nearest neighbor algorithm (KNN) and the standard active and passive GPM products (Fig.~\ref{fig06}). To quantify statistical agreements between the results of the algorithm and those of the REF maps, we calculate the annual probability of detection, false alarm, and the Heidke skill score \citep{Doswell1990} for the presented results in Figs.~\ref{fig07},\ref{fig08} and \ref{fig09}. We also compare the algorithm outputs with the precipitation phase products of the Multi-Radar/Multi-Sensor System (MRMS) on a seasonal basis (Figs.~\ref{fig10} and \ref{fig11}). Finally, some results are presented at a storm-scale to demonstrate the detection capabilities of the algorithm for a few precipitation events that are coincidentally captured by the DPR, high-resolution ground-based radars (Figs.~\ref{fig12} and \ref{fig13}), and simulated by the Weather Research and Forecasting (WRF) model (Fig.~\ref{fig14}) during the Olympic Mountain Experiment \citep[OLYMPEx,][]{Houze2017}.

\subsection{Global Retrievals}

\begin{figure*}[!ht]
 \centering
 {\includegraphics[width=0.9\textwidth]{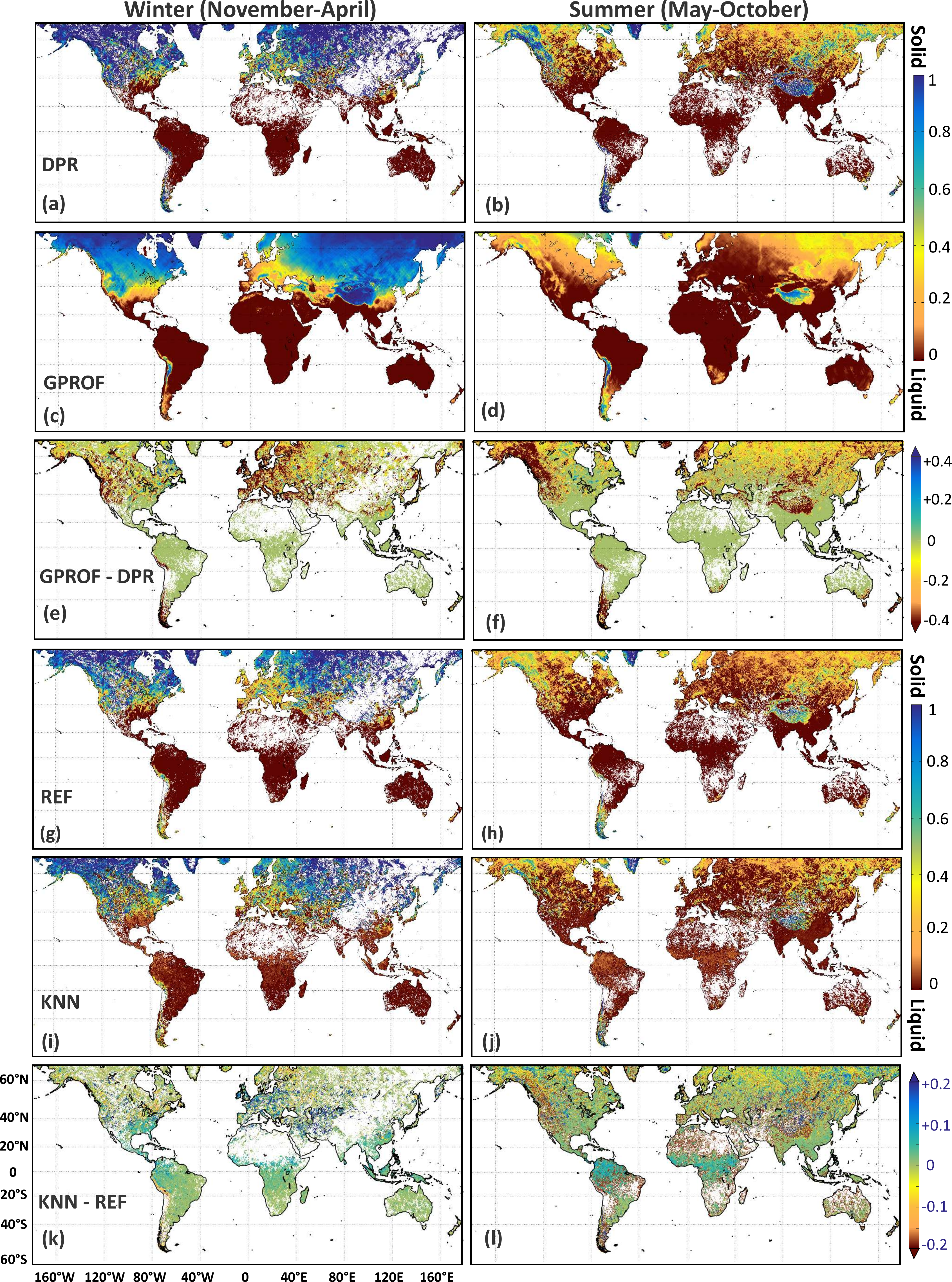}}\vspace{-2mm}
\caption{Seasonal probability of the precipitation phase change. The average phase of the DPR (a,\,b), GPROF (c,\,d), REF (merged) (g,\,h) and the KNN algorithm (i,\,j) as well as the differences between the DPR and GPROF (e,\,f) and the REF and KNN products (k,\,l). The differences are shown where both products detect the precipitation occurrence.}\label{fig06} 
\end{figure*}

The seasonal average of the quasi-global maps of precipitation phase are presented in Fig.~\ref{fig06}, for the inner-swath data products by the DPR, GPROF, KNN, and REF. The results are shown as a probability continuum of phase transition from the liquid ($0$) to solid ($1$), at the grid resolution 0.1-degree. These results are mapped where the precipitation is detected only by the DPR for two seasons. The seasons are defined as summer (June-October 2015 and May 2016) and winter (November 2015 to April 2016) of the Northern Hemisphere. 

Overall, since the phase of the passive product is dictated by the reanalysis data, the results mostly follow the climatology patterns of near-surface wet-bulb temperature and are smoother than those of the active product (Fig.~\ref{fig06}\,a-d). The smoothness of the GPROF retrievals could also be due to its ability in retrieving the light precipitation with intensities below the minimum detectable rate by the DPR ($<0.2$~\si{mm.h^{-1}}), as the GPROF also uses precipitation data from MRMS ground-based radar in its a priori database to increase the retrieval sensitivity to snowfall. Comparison of the official passive and active products remains outside the scope of this research; however, there seem to be notable differences in the spatial patterns of precipitation phases in these products. The difference in the source of ancillary data could be a reason for the observed discrepancies, which largely exist over mountainous terrains such as the Andes, Tibetan highlands, Rockies, Scandinavian mountains, Alps, and Zagros Mountains (Fig.~\ref{fig06}\,e,\,f)\,---\,where precipitation is mostly triggered by topographic features. 

The observed differences are not surprising because of complications in both active and passive retrievals due to reduced ice scattering in shallow orographic lifting, heterogeneity of surface roughness, and radiometric complexity of high -elevation snow and ice cover \citep{Tian2010}. The phase discrepancies also seem to be larger when it comes to identifying precipitation phase in the summer. For example, over the Tibetan highlands, the active products classify most of the summer precipitation as snowfall while the passive product results in more liquid precipitation, especially over the Hengduan Mountains in southeast China.

\begin{figure*}[!ht]
 \centering
 {\includegraphics[width=\textwidth]{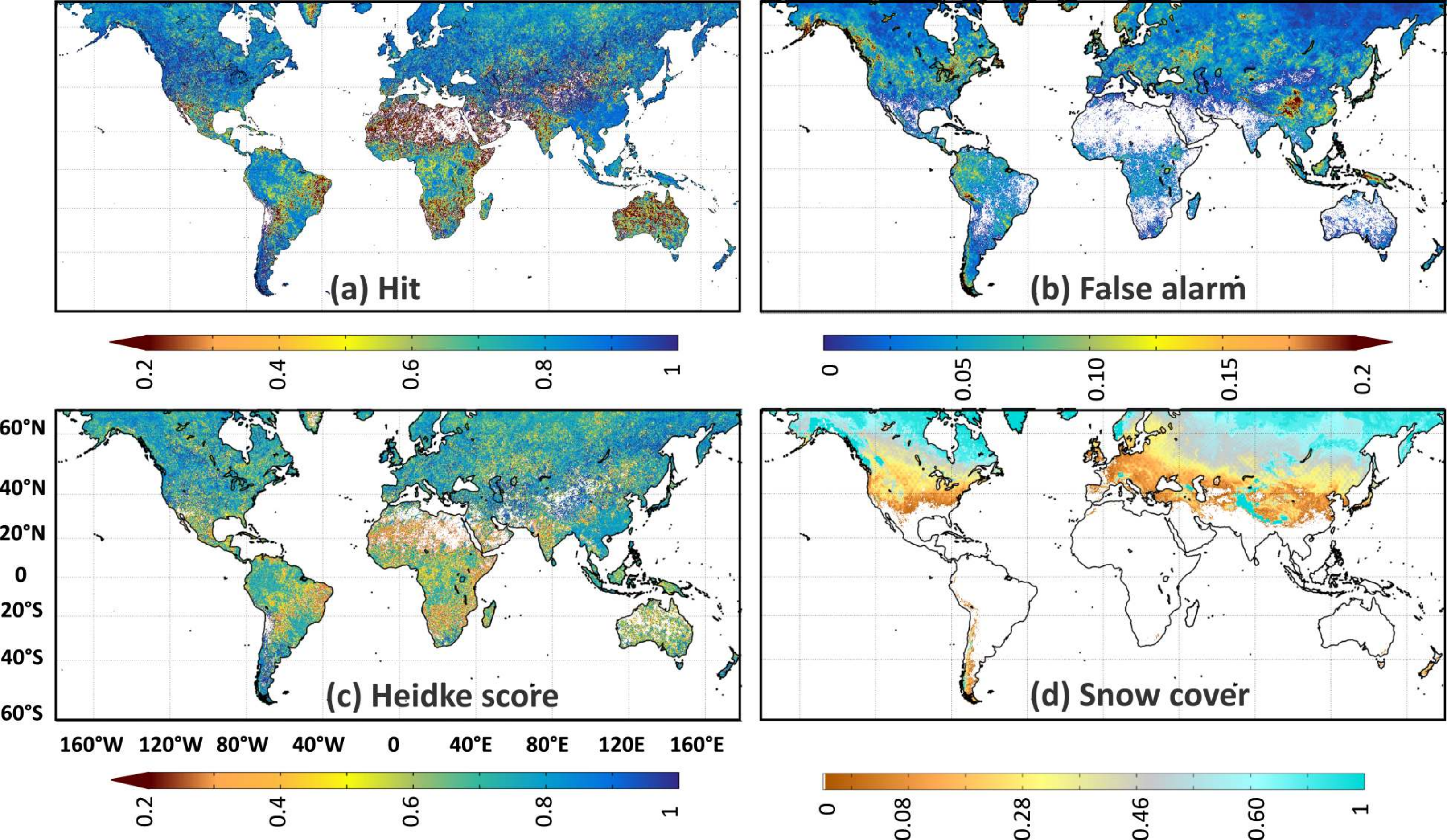}}\vspace{-4mm}
\caption{The mean annual map of the probability of hit (a), probability of false alarm (b), and Heidke score (c) obtained by comparing the pixel-level results of the KNN algorithm with the REF product for the detection of precipitation occurrence. The map of snow-cover fraction (d) is also obtained from the MODIS data \citep[MOD10A1][]{Hall2002} coincident with GPM inner-swath overpasses from June 2015 to May 2016.}\label{fig07} 
\end{figure*}

Fig.~\ref{fig06}\,i,\,j shows the results of the KNN algorithm in summer and winter and compare them with the REF map (Fig.~\ref{fig06}\,k,\,l). Overall, we observe a good agreement between the KNN outputs and the REF target precipitation product. The differences are more pronounced in the summer than the winter and mostly accumulated over the mountainous and dense vegetation regions (Fig.~\ref{fig06}\,k,\,l). For example, we observe that, in the summer, the detection probability of solid and mixed phases are negatively biased ($\sim -12\%$) over the Rockies and the Andes. However, in winter, this probability is positively biased over small parts of the Scandinavian mountains in northern Europe ($\sim+15\%$). Some of these mountainous biases are mainly attributed to the false detection of precipitation occurrence rather than its phase (Fig.~\ref{fig07}\,b). Additionally, over the tropical forests, the algorithm falsely detects some mixed precipitation phases. Over dense vegetative surfaces the microwave polarization signal becomes very weak \citep{Prigent1997} due to incoherent vegetation scattering. The lack of a pronounced polarization signal could be the main reason for the reduced discriminatory power of the KNN approach that relies on the Euclidean distance as a similarity metric.

Visual inspection of the global maps shows a good spatial and seasonal agreement between the KNN and REF. The proximity of these two products at the global scale is quantified by three measures including the Spearman\textquotesingle s correlation ($\rho$), the Root Mean Squared Error (RMSE), and the Kullback-Leibler divergence ($KL$) in Fig.~\ref{tab01}. The $KL$ divergence $KL({P}\parallel{Q})=\sum_{i=1}^{n} \dfrac{P(i)}{Q(i)}$ is a non-symmetric and non-negative measure that captures the proximity of two probability distributions $P$ and $Q$ and is zero when they are identical. To compute the $KL$-divergence, between the probability histograms of the REF ($P$) and KNN outputs ($Q$), we discretize $P$ and $Q$ with $n = 20$ probability intervals. The RMSE and $KL$ values are normalized between 0 and 1 for interpretation convenience. As is evident, the correlation between the KNN and REF products is around 0.89 to 0.91 in winter and summer, indicating that the algorithm is not excessively sensitive to the seasonal changes in land surface radiometric properties. The normalized RMSE also remains below 14\% in both seasons. We see that the $KL$ values slightly increase from winter (0.06) to summer (0.10), which indicates that, on average, the KNN method may exhibit improved detection skills when the extent of the global snow cover is larger in winter than summer.

\begin{figure*}[tbh!]
 \centering
 {\includegraphics[width=\textwidth]{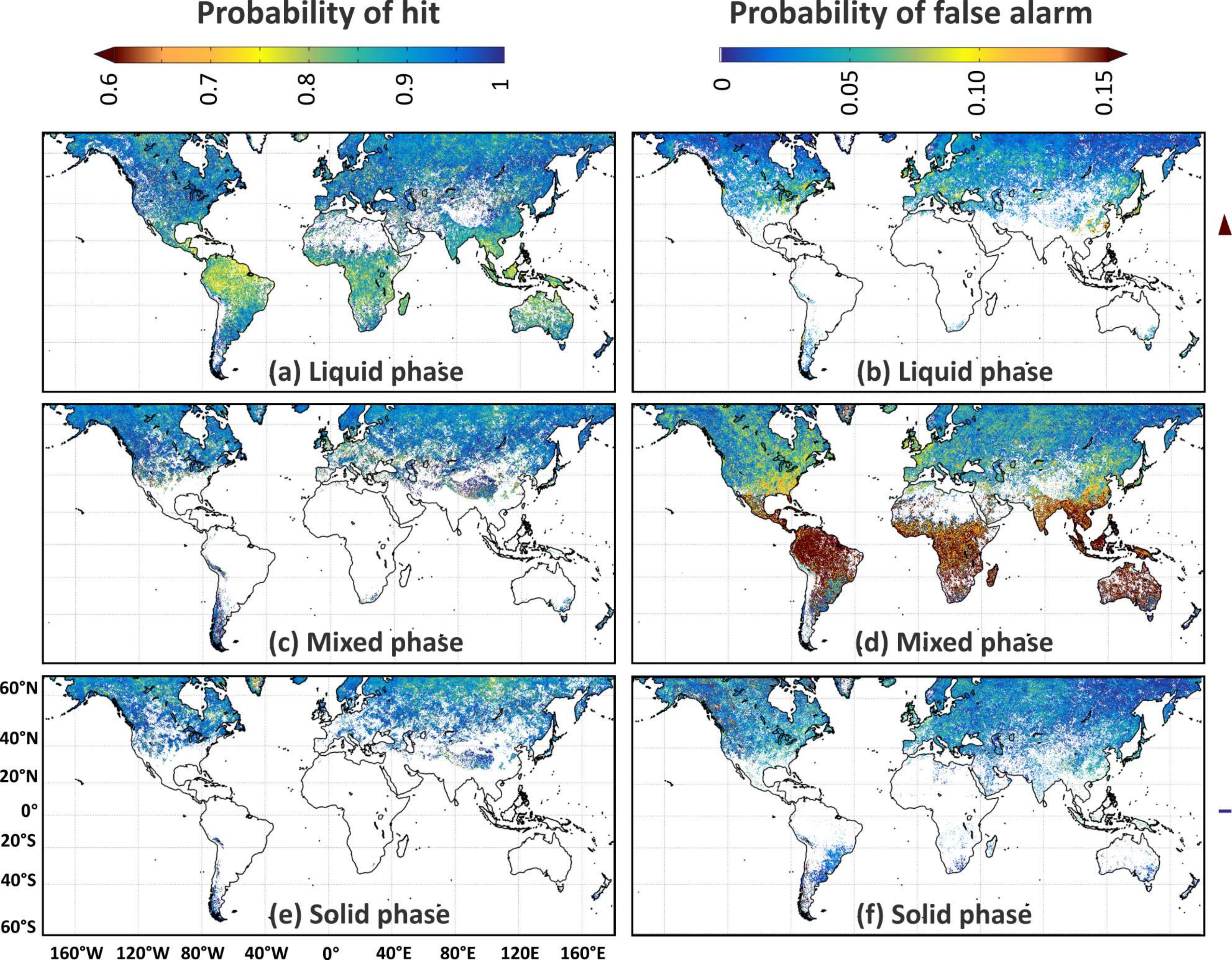}}
\caption{The mean annual map of the probability of hit and false alarm by the KNN algorithm for the detection of the liquid phase (a,\,b), mixed phase (c,\,d) and solid phase (e,\,f). The results are obtained for all GPM inner-swath overpasses from June 2015 to May 2016.}\label{fig08} 
\end{figure*}

\begin{figure*}[!ht]
 \centering
 {\includegraphics[width=17pc]{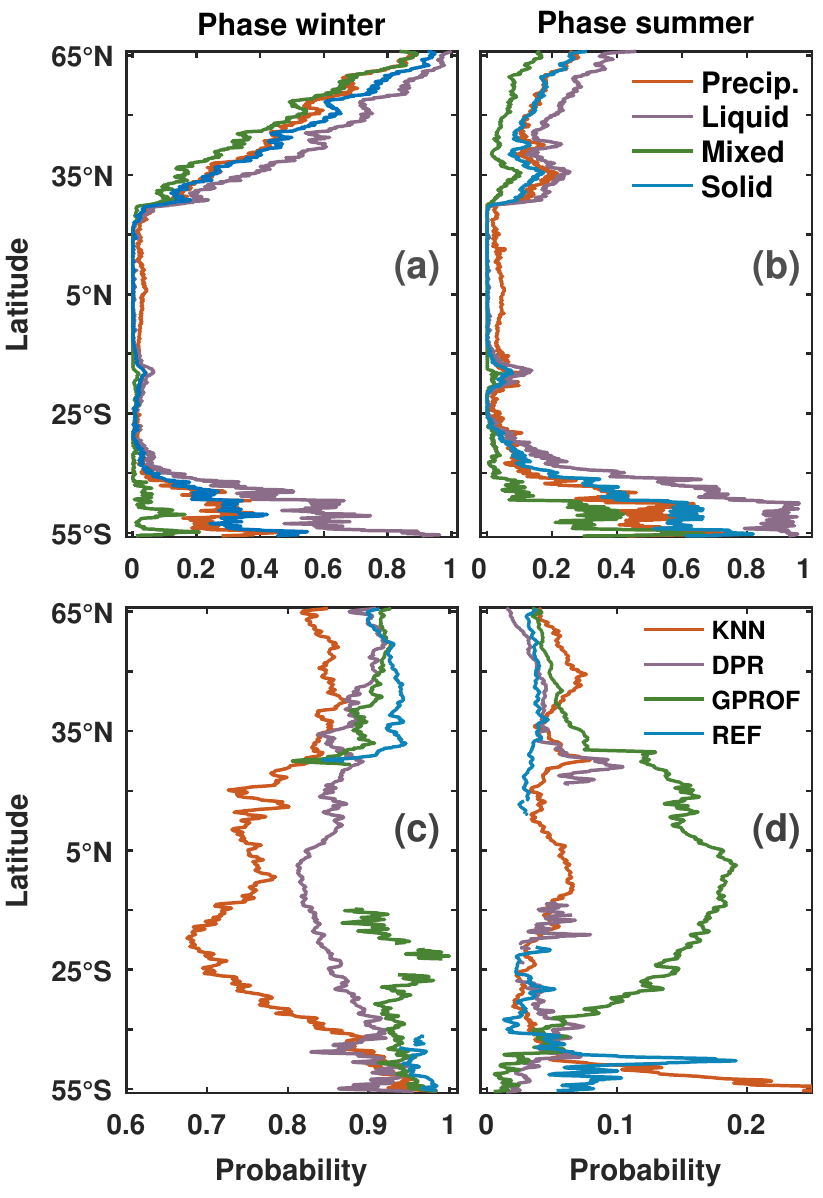}}\vspace{-4mm}
\caption{Zonal mean values of the probability of precipitation phase change from liquid ($p=0$) to solid ($p=1$) by the KNN, DPR, GPROF, and REF products in (a) winter (November-April) and (b) summer (May-October). Zonal mean values of (c) probability of hit and (d) false alarm for the detection of the precipitation occurrence and its phase change by comparing the KNN results with the REF product.}\label{fig09} 
\end{figure*}

To further reveal the error structure of the instantaneous pixel-level retrievals, we used three statistical measures including the probability of hit, probability of false alarm, and the Heidke skill score $HSS = \frac{2(ad-bc)}{(a+c)(c+d)+(a+b)(b+d)}$ \citep{Doswell1990}, which ranges from a no skill ($-\infty$) to a perfect skill (1). Recall that $a$ is the number of correctly detected events, $c$ is the number of missed events, $b$ is the number of false detection, and $d$ is the number of correct rejection. To have an adequate number of samples, these quality measures are calculated using the entire validation period from June 2015 to May 2016 (Table~\ref{tab02} and Figs~\ref{fig07}~ and~\ref{fig08}).

\begin{table}[hbt]
\caption{Quality metrics obtained by comparing the annual probability of phase transition between the KNN results and the reference product (REF). Shown statistics are the normalized Root Mean Squared Difference (RMSD), Spearman\textquotesingle s correlation ($\rho$), and the Kullback-Leibler divergence ($KL$).}
\begin{tabular*}{\hsize}{@{\extracolsep\fill}lccccc@{}}
\toprule
Metrics & $\rho$ & RMSE & $KL(\Delta{p} =0.05$)\\
\midrule
\ Winter (November-April)  & 0.91   & 0.12 & 0.06\\
\ Summer (May-October)  & 0.89   & 0.14 & 0.10\\
\hline 
\end{tabular*}\label{tab01}
\end{table}

The annual maps of the probability of hit, false alarm, and HSS score are used to evaluate the detection skill of the KNN approach against the DPR as a reference (Fig.~\ref{fig07}). The probability of hit over the snow-covered regions is relatively high. The reason is that the presence of snow on the ground reduces the surface emission, which could lead to better detection of the precipitation emission signal (Fig.~\ref{fig03})\,---\,similar to radiometrically cold ocean surfaces. The low detection rates are mostly over the areas where the DPR has a low sampling rate. Thus, lack of skills in these regions could be partly due to lack of samples in the database. A high probability of false alarm ($\sim$\,0.2) is seen over some mountainous regions such as the Tibetan highlands and the Western Rockies. The false detection, mostly in liquid phase, gives rise to negative biases in detecting frozen and mixed precipitation (Fig.~\ref{fig06}\,l). High ($\sim$\,0.80), medium ($\sim$\,0.66) and low values ($\sim$\,0.50) of HSS score are observed over the snow cover, tropical forests, and under-sampled deserts such as Sahara, respectively (Fig.~\ref{fig07}\,c).

\begin{table*}[bth]
\caption{The annual probability of hit and false alarm for the KNN results over different land surface types $\left\lbrace{\mathcal{L}} \right\rbrace_{s=1}^{3}$ and detection classes $\left\lbrace{\mathcal{D}} \right\rbrace_{i=1}^{4}$. Here, $s = 1$ to 3 represents the ground, wet, and dry snow-covered surfaces, $i = 1$ denotes the detection of precipitation occurrence, and $i = 2$ to 4 represents the detection of liquid, mixed, and solid phase, respectively. The results over the dry snow cover ($\mathcal{L}_{3}$) are further stratified based on the annual percentage of the snow.}
\vspace*{4mm}
\label{tab02}
{\renewcommand{\arraystretch}{1.4}
\begin{tabular*}{\hsize}{@{\extracolsep\fill}ccccccccc@{}}
\multicolumn{9}{c}{\textbf{Probability of hit}}\\
\toprule
& \multicolumn{3}{c}{Land surface} & \multicolumn{5}{c}{Annual percentage of dry snow cover ($\mathcal{L}_{3}$)}\\
\hline
 & $\mathcal{L}_{1}$ & $\mathcal{L}_{2}$ & $\mathcal{L}_{3}$ &{$0-0.10$} & {$0.10-0.25$} &{$0.25-0.45$} &{$0.45-0.70$} &{$0.70-1.00$}\\
\cmidrule(lr){2-4} \cmidrule(lr){5-9}
$\mathcal{D}_{1}$ & 0.75 & 0.78 & 0.86 & 0.77 & 0.84 & 0.86 & 0.84 & 0.87\\
$\mathcal{D}_{2}$ & 0.82 & 0.83 & 0.90 & 0.85 & 0.89 & 0.90 & 0.91 & 0.92\\
$\mathcal{D}_{3}$ & 0.86 & 0.89 & 0.92 & 0.88 & 0.90 & 0.93 & 0.92 & 0.92\\
$\mathcal{D}_{4}$ & 0.86 & 0.86 & 0.94 & 0.88 & 0.93 & 0.94 & 0.96 & 0.95\\
\addlinespace
\addlinespace
\multicolumn{9}{c}{\textbf{Probability of false alarm}}\\
\toprule
& \multicolumn{3}{c}{Land surface} & \multicolumn{5}{c}{Annual percentage of dry snow cover ($\mathcal{L}_{3}$)}\\
\midrule
 & $\mathcal{L}_{1}$ & $\mathcal{L}_{2}$ & $\mathcal{L}_{3}$ &{$0-0.10$} & {$0.10-0.25$} &{$0.25-0.45$} &{$0.45-0.70$} &{$0.70-1.00$}\\
\cmidrule(lr){2-4} \cmidrule(lr){5-9}
\addlinespace
$\mathcal{D}_{1}$ & 0.06 & 0.09 & 0.11 & 0.09 & 0.08 & 0.13 & 0.09 & 0.11\\
$\mathcal{D}_{2}$ & 0.10 & 0.05 & 0.04 & 0.07 & 0.05 & 0.06 & 0.04 & 0.04\\
$\mathcal{D}_{3}$ & 0.09 & 0.08 & 0.05 & 0.08 & 0.05 & 0.07 & 0.06 & 0.05\\
$\mathcal{D}_{4}$ & 0.08 & 0.05 & 0.04 & 0.08 & 0.05 & 0.05 & 0.04 & 0.04\\
\bottomrule
\end{tabular*}}
\end{table*}

The conditional probability of hit and false alarm are calculated for liquid, mixed, and solid phases (Fig.~\ref{fig08}), with respect to the REF product. For separating the errors of the precipitation and phase detection, the probabilities are obtained assuming that the precipitation is correctly detected by the KNN algorithm. Similar to the precipitation detection, the algorithm displays improved phase detection capabilities over snow-covered surfaces (Fig.~\ref{fig08}). The probability of hit for the liquid, mixed, and the solid phase is mostly greater than 0.85 and reaches 0.95 over the high-altitudes of North America. However, we observe a relatively lower detection rate of around 0.74 for liquid precipitation over the tropical and subtropical regions such as the rainforest of Amazonian and Central Africa. The results show that the low probability of detection for the liquid phase is mostly because the algorithm detects some false mixed phase precipitation (Fig.~\ref{fig08}\,d). We speculate that this error could be partly attributed to the reduced skill of the algorithm over vegetated surfaces. The reduced detection skill of the algorithm could also be partly due to warm rain occurrences over the heterogeneous land surface of tropical and subtropical regions where cloud ice scattering is not significant.

To understand the reasons for improved retrievals over snow-covered surfaces, the averaged values of the probability of hit and false alarm are stratified based on precipitation occurrence $(\mathcal{D}_{1})$ at liquid $(\mathcal{D}_{2})$, mixed $({\mathcal{D}}_{3})$ and the solid $(\mathcal{D}_{4})$ phase over different land surface types $\left\lbrace{\mathcal{L}} \right\rbrace_{s=1}^{3}$, where $s=1$ to 3 denotes the ground, wet, and dry snow cover (Table~\ref{tab02}). The probability of precipitation detection increases by almost 11\% from the ground to the dry snow cover, and 3\% from wet to dry snow. An increase of 8 to 11\% is also observed in the probability of hit in detection of solid and liquid phase over dry snow, where the largest detection rate of 94\% is obtained for the snowfall. The results show that the probability of false alarm also increases in detection of precipitation occurrence over snow cover, whereas it decreases when it comes to the detection of its phase. Because, once precipitation is detected, due to significant differences between the signatures of rain and snowfall, the probability of false alarm is markedly reduced. Table~\ref{tab02} quantifies the dependency of the probability of hit and false alarm on the annual percentage of the dry snow cover. For precipitation detection, the probability of hit increases by about 10\% as the annual percentage of dry snow increases from zero to more than 70\%, while the probability of false alarm increases between 2--4\%. As is evident, for precipitation phase detection, both probabilities show improvements of around 4\%.

\FloatBarrier

\subsection{Comparison with the ground-based radar}
\subsubsection{Comparison with Multi-Radar/Multi-Sensor System}
To further evaluate the performance of the KNN algorithm, we compare its outputs against a precipitation product derived from the Multi-Radar/Multi-Sensor System (MRMS) \citep{Zhang2011, Zhang2016}. MRMS mosaics three-dimensional volume scan observations from 146 S-band dual-polarization Doppler Weather Surveillance Radar-1988 (WSR-88D) and 31 C-band single polarization Canadian radars. The product optimally integrates the radar observations with simulations of atmospheric models as well as hourly gauge data to produce seamless precipitation rate and phase estimates over the CONUS, at spatial resolution of 1\,km at every 2\,min. The MRMS products are further quality-controlled and gauge-adjusted at fine scale following the procedure described in \citet{kirstetter2012} to derive a consistent and high quality surface precipitation.

To determine the precipitation phase, MRMS uses thresholds on the wet and dry bulb temperatures. Specifically, the precipitation is labeled as snowfall when the radar reflectivity exceeds 5\,dBZ, the surface temperature is below \SI{2}{\celsius}, and the surface wet bulb temperature is below \SI{0}{\celsius} \citep{Zhang2016}. Thus the MRMS rain-snow delineation is subject to similar uncertainties as in the passive GPM data \citep{Chen2016}. However, the uncertainties in detecting precipitation are significantly lower than the satellite data because of the higher sensitivity and resolution of the ground-based radar observations, especially over landscapes with no significant orographic features \citep{kirstetter2012}. To compare with the outputs of the KNN algorithm, a reference surface precipitation is derived by mapping the high-resolution MRMS data onto, and then averaging over, the nearest DPR grids \citep[see][]{kirstetter2012,Kirstetter2014}. 

\begin{figure}[!ht]
\centerline{\includegraphics[width=\textwidth]{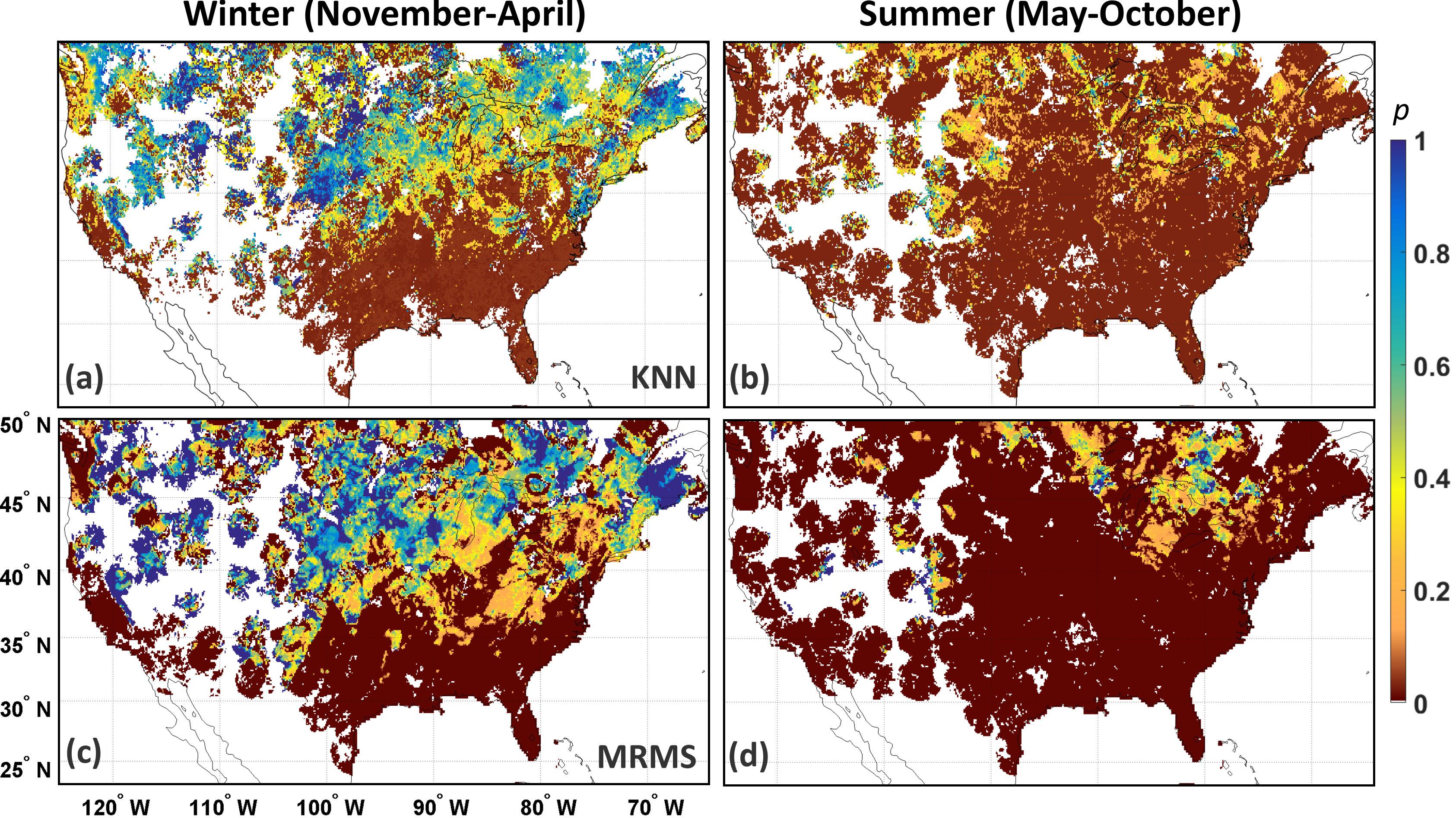}}
\caption{Mean seasonal maps of the probability of precipitation phase change from liquid ($p=0$) to solid ($p=1$) for KNN in winter (a) and summer (b), and for the MRMS in winter (c) and summer (d), from June 2015 to May 2016}\label{fig10} 
\end{figure}

\begin{table}[h!]
\caption{Quality metrics obtained by comparing the annual probability of phase transition between the KNN retrievals and MRMS observations. Shown statistics are the normalized Root Mean Squared Difference (RMSD), Spearman\textquotesingle s correlation ($\rho$), and the Kullback-Leibler divergence ($KL$).}
\begin{tabular*}{\hsize}{@{\extracolsep\fill}lccccc@{}}
\toprule
Metrics & $\rho$ & RMSE & $KL(\Delta{p} =0.05$)\\
\midrule
\ Winter (November-April)   & 0.72   & 0.29 & 0.27\\
\ Summer (May-October)   & 0.78   & 0.21 & 0.15\\
\hline
\end{tabular*}\label{tab03}
\end{table}
\FloatBarrier

\begin{figure}[htb!]
\centering
{\includegraphics[width=0.6\textwidth]{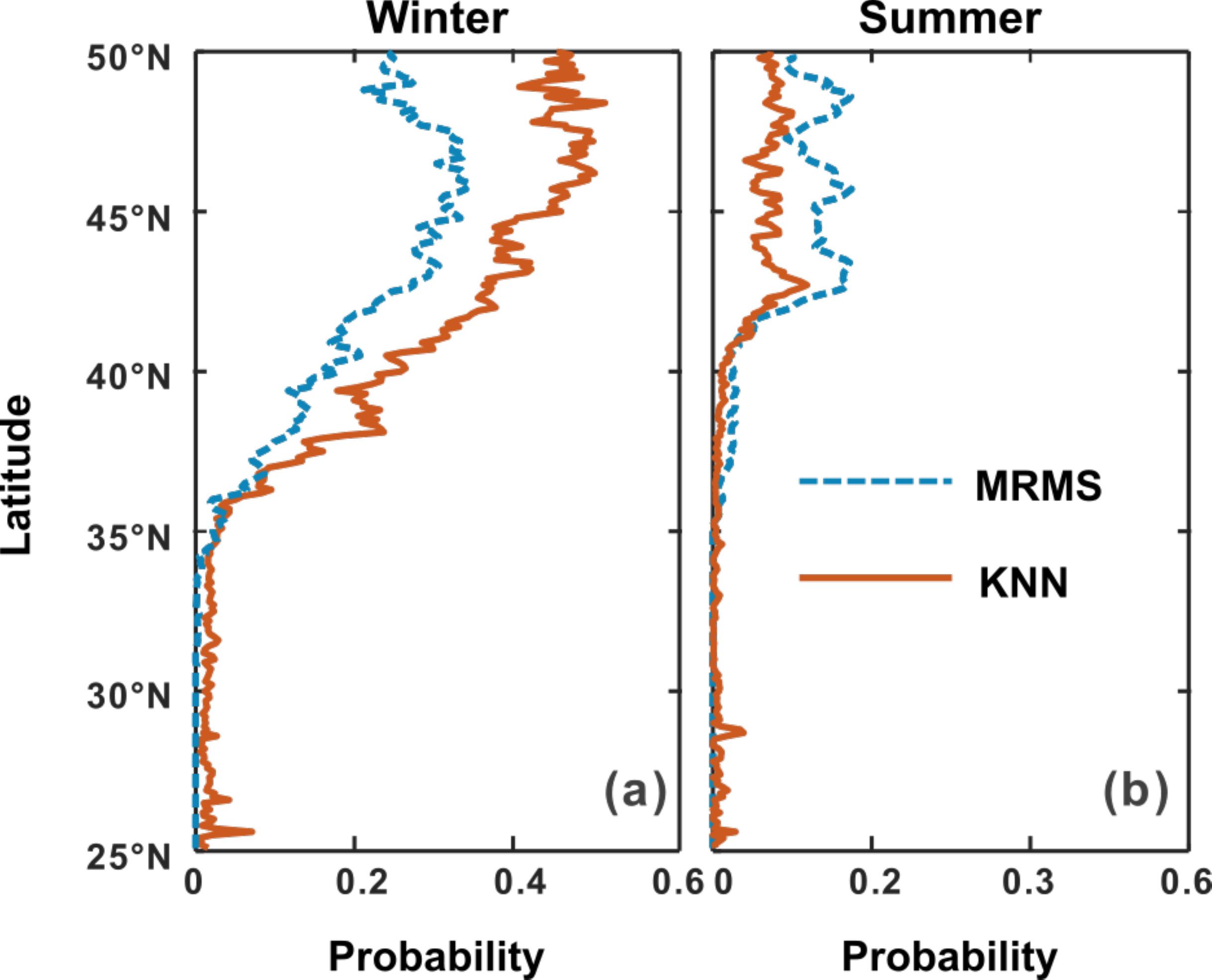}} \vspace{-4mm}
\caption{The zonal mean of the probability of precipitation phase change from liquid ($p=0$) to solid ($p=1$) by the KNN and MRMS products in (a) winter (November-April) and (b) summer (May-October)\,---\,from June 2015 to May 2016.}\label{fig11} 
\end{figure}

Fig.~\ref{fig10} shows that the spatial variations of the probability of phase change in the KNN and MRMS are consistent in the winter and summer seasons. The calculated values of $KL$-divergence between KNN and MRMS are 0.27 and 0.15 in winter and summer, respectively. The values of other calculated similarity metrics (i.e., $\rho$ and RMSE) are also deteriorated from summer to winter (Table~\ref{tab03}). These results indicate that even though the KNN shows improved wintertime detection of precipitation compared to those in summertime when compared with the REF product (Table~\ref{tab03}), the intrinsic error between the satellite and ground-based data is still much larger than the satellite retrieval error, especially in the winter. The zonal mean of the phase transition probabilities (Fig.~\ref{fig11}) indicates more similarities at lower latitudes ($<40$\si{\degree} N), where the uncertainty of precipitation phase change is lower or remains close to zero. At higher latitudes, KNN generates a higher (lower) probability of snowfall occurrence relative to the MRMS in winter (summer). In particular, larger departures occur at latitudes higher than \SI{37}{\degree}~N in winter and \SI{43}{\degree}~N in summer, where the ground is usually covered with snow.

\begin{figure}[htb!]
\centering
{\includegraphics[width=\textwidth]{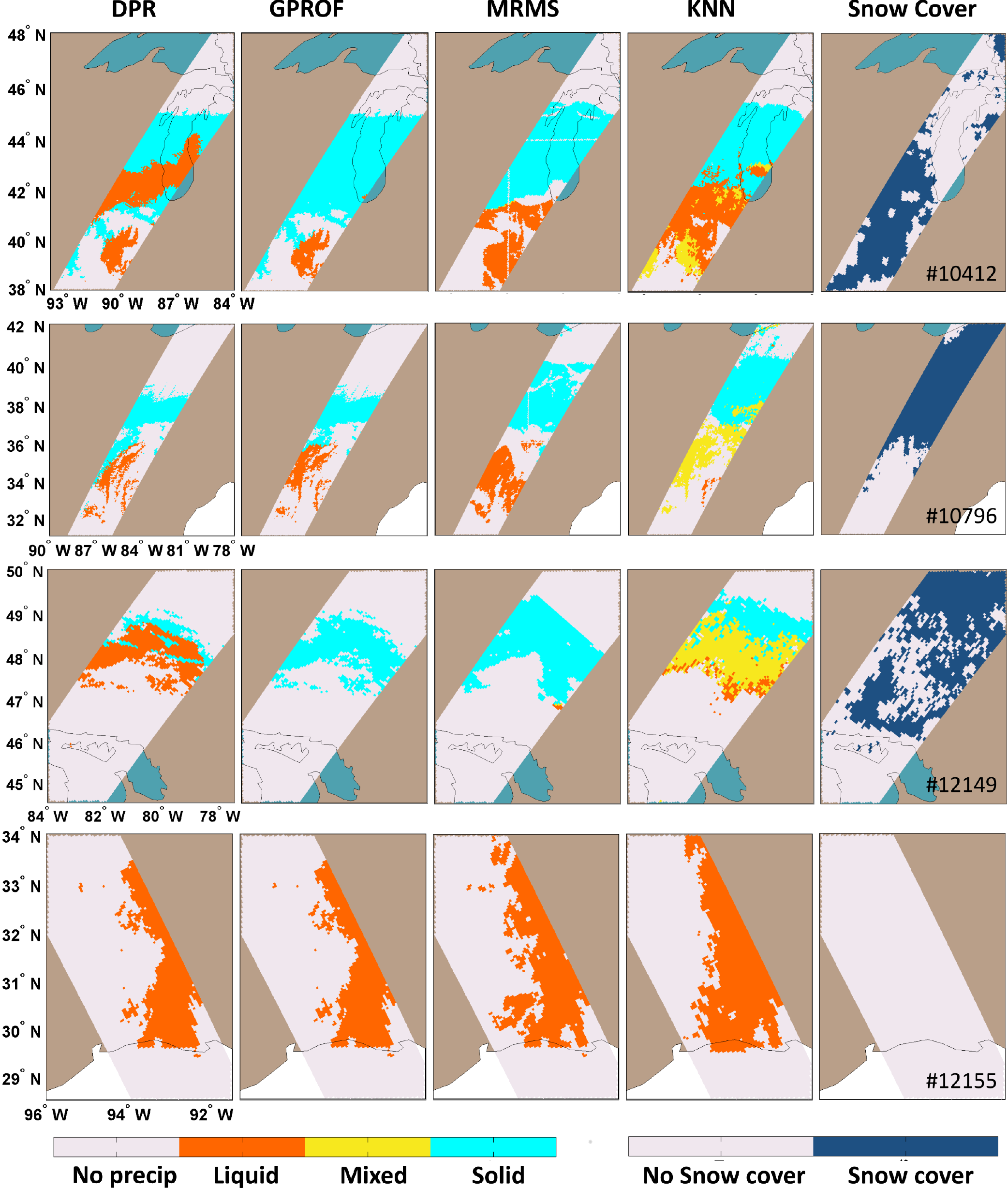}}\vspace{-5mm}
\caption{Orbital-level precipitation phase detection from the KNN, DPR, GPROF, and MRMS for a few GPM overpasses including $\#10412$ on 2015/12/28 (top row), $\#10796$ on 2016/01/22 (second row), $\#12149$ on 2016/04/18 (third row), and $\#12155$ on 2016/04/18 (bottom row).}\label{fig12} 
\end{figure}

Fig.~\ref{fig12} shows four different satellite overpasses that capture large storms with distinguishable spatial phase change. Overall, the KNN approach is skillful in capturing the occurrence and phase of the near-surface precipitation. As is evident, in case of a single-phase precipitation event (e.g. orbit $\#12155$), the KNN can accurately detect the extent of the storm, especially when a large part of the storm is in liquid form. However, when the several phases coexist within the storm (e.g. orbits $\#10412$, $\#10796$ and $\#12149$), discrepancies arise between the satellite active/passive products and the MRMS data. The produced mixed phase by the KNN retrieval reflects the uncertainty between the satellite active/passive retrievals where a freezing point is likely to occur in the DPR ground clutter zone. For example, the storm on the northern shores of Lake Huron (orbits $\#12149$) is well detected in terms of its spatial extent. The phase detection in the GPM passive product (GPROF) and the MRMS products is consistent since both products rely significantly on the wet-bulb temperature data. However, the DPR product differs significantly from other products and produces more liquid phase over the southern edge of the storm. As is evident, the KNN retrievals capture this discrepancy through a mixed phase detection.

\begin{figure*}[!htb]
 \centering
 {\includegraphics[width=\textwidth]{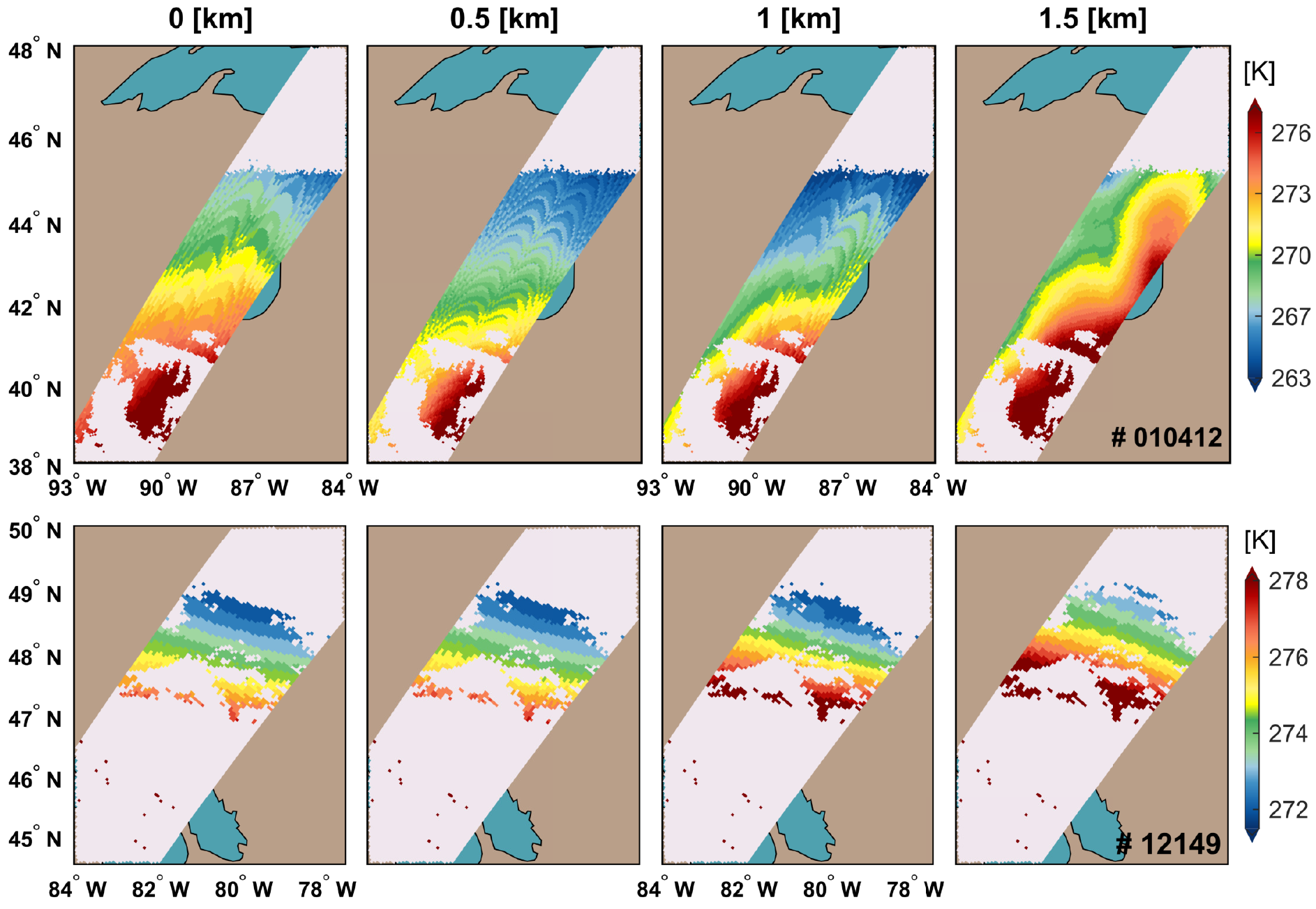}}
\caption{Inversion of the air temperature at orbit $\#10412$ on 2015/12/28 (top row) and orbit $\#12149$ on 2016/04/18 (second row). The data (2A-DPRENV) are presented at four ranges from 0 (surface) to \SI{1.5}{\km}.}
\label{fig13} 
\end{figure*}

It is surprising that in orbits $\#10412$ and $\#12149$ \mbox{(Fig.~\ref{fig12})}, the DPR reports the phase as liquid where the GPROF classifies the phase largely as solid since the discrepancy is often in the other direction. Based on the atmospheric temperature profile derived from environmental ancillary data (2A-DPRENV) used in the active retrieval algorithm, we conclude that the there is a temperature inversion when the storm is happening (see Fig.~\ref{fig13}). In this case, liquid precipitation can refreeze near the surface and may not be captured by the DPR.

\subsubsection{Comparison with the WRF simulations during the OLYMPEx}
\begin{figure*}[!htbp]
 \centering
 {\includegraphics[width=\textwidth]{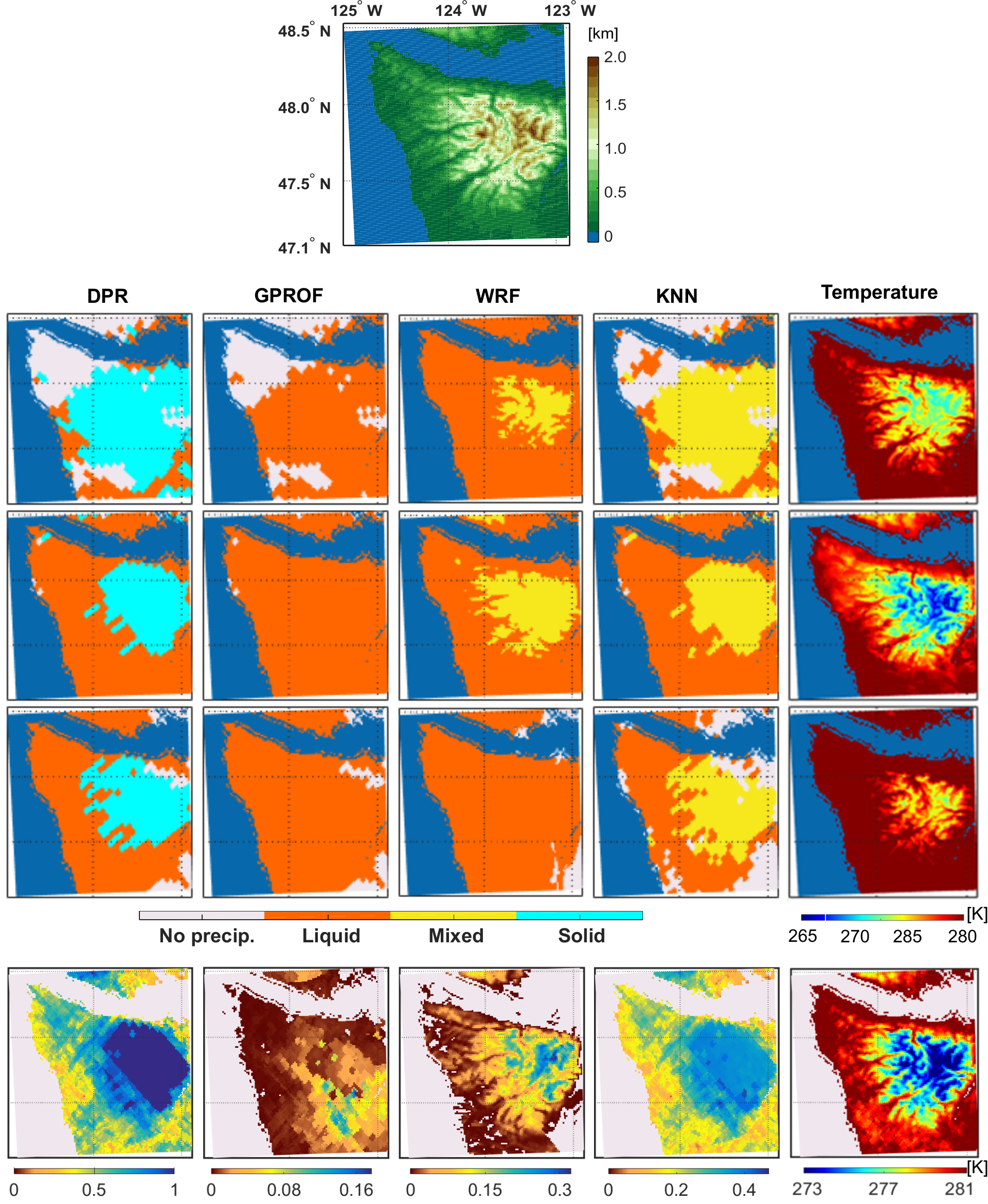}} \vspace{-4mm}
\caption{The digital elevation map (DEM) of the Olympic Mountains (top panel), and the precipitation phase by the DPR, GPROF, WRF, and KNN for orbit numbers $\#9722$ (2015/11/14, second row), $\#9773$ (2015/11/17, third row), and $\#10019$ (2015/12/03, fourth row). The bottom panel is the average probability of phase for 117 GPM inner-swath overpasses from November 1 to December 23, 2015. The last column shows the 2-meter air temperature from the WRF simulations.}\label{fig14} 
\end{figure*}

The MRMS data lacks coverage over mountainous regions, thus we need a venue with rich ground-based observations for further evaluation of the presented approach. There is a wealth of orographic precipitation data during the GPM Olympic Mountains Experiment \citep[OLYMPEx,][]{Houze2017} from November 1 to December 23, 2015. The Olympic Mountains are located in the northwestern corner of the Washington State, United States (Fig.~\ref{fig14}) with a dominant orographic precipitation regime. This regime is a result of the abrupt uplift of moisture-laden southwest airflow coming from the mid-latitude baroclinic storm systems. A few high-elevation snow and precipitation gauges were used during the OLYMPEx field campaign. However, the coarse temporal resolution of the DPR (i.e., 117 partial overpasses), relative to the 56-day duration of OLYMPEx hamper their use for our purpose. Therefore, we choose the outputs of a high-resolution (1.33 km) hourly WRF simulation by the Northwest Modeling Consortium over the Olympic Mountains \citep{Mass2003}.

\citet{currier2017independent} used the microphysical scheme of WRF to estimate precipitation phase and showed that the results are relatively unbiased when compared with the OLYMPEx ground-based observations. The data is available from November 2015 to May 2016 and contains almost 117 full or partial overlaps with DPR overpasses. First, the DPR retrievals are spatially resampled to match the 1.33 km WRF outputs. Then, the hourly outputs of the WRF are interpolated to match the scanning time of the DPR. To convert the interpolated WRF outputs to discrete precipitation phase, we follow a simple rule. If the ratio of reported snowfall to rainfall intensity is higher (lower) than 0.66 (0.33), then the precipitation is considered as solid (liquid) phase; otherwise, it is labeled as mixed. 

Fig.~\ref{fig14} illustrates the precipitation phase for the DPR, GPROF, KNN, and the WRF for three GPM orbits ($\#9722$, $\#09773$, $\#10019$). We observe that at high-elevation regions, the KNN detects mixed phase over the areas that exhibit phase discrepancies between the GPROF and DPR. We see that these KNN results are in a good agreement with the WRF simulations. However, it is important to note that the precipitation phase partitioning in the WRF outputs is based on cloud microphysical parameters in the atmospheric boundary layer, and thus its mixed-phase precipitation is physically different than the defined mixed-phase category in KNN retrievals.

We calculate and compare the average phase outputs of the DPR, GPROF, KNN, and WRF data for all 117 coincident DPR overpasses. We fond that compared with the average phase probability of WRF, the KNN precipitation phase is positively biased by about 28\% (i.e., KNN captures more solid phase than WRF, Fig.~\ref{fig14}). However, this bias is about 31\% at elevations above 800 meters, while reduced to about ${24\%}$ for lower elevations. Additionally, the results show that over areas with elevations higher than 800 meters, the KNN phase bias is significantly smaller compared to both DPR (positive bias $\sim{48\%}$) and GPROF (negative bias $\sim{56\%}$). At elevations below 800 meters, the KNN is less biased than the positively biased DPR ($\sim{41\%}$); however, about 9\% more biased than GPROF with a negative bias $\sim{19\%}$. Overall, these results indicate that even though the KNN phase detection is consistent with the satellite products, there are notable discrepancies with the WRF simulations over the mountainous regions, which need further investigation.
\FloatBarrier
\section{Summary and Discussion}

We proposed a Bayesian algorithm for detection of precipitation occurrence and phase from satellite observations, with particular emphasis on snowfall detection over snow cover. The algorithm relies on a nested \textit{k}-nearest neighbor (KNN) search and probabilistic vote rules for detection of precipitation occurrence and its phase. The a priori database in the algorithm contains collocated GMI brightness temperatures (10.65 to 183 GHz) and DPR precipitation data that were stratified based on snow-cover retrievals from the MODIS sensor on board the Terra satellite. The precipitation phase data from the GPM passive and active products were combined to provide a reference database for testing the skill of the algorithm. 

The results demonstrated that the weighted Euclidean distance can be used as a similarity metric for precipitation phase detection in a Bayesian setting, with improved results over snow-covered surfaces. We demonstrated that the KNN is able to identify precipitation phase with minimal dependency on ancillary data, such as the near-surface air temperature and moisture. The results showed that the global probability of hit for detection of solid precipitation over dry snow cover could reach up to $\sim$\,94\%. However, the detection skill of the algorithm is decreased over regions with dense vegetation due to reduced polarization signal. A larger phase discrepancy was found when the KNN results were compared with the ground-based precipitation phase, which remains to be addressed in future research.

It is important to emphasize that we have used V04 GPM official products. We expect to see less discrepancies between the GPM retrievals and the ground-based phase products in the following versions, because the latest version of the GPROF phase detection algorithm benefits from the longer GPM Radar/Radiometer joint records and the new DPR algorithm relies on an improved parameterization of ice microphysics.

Linking the algorithm with physical or observational databases that contain additional information on snow-cover physical properties (e.g., snow thickness, density, and liquid water content) and vegetation density can be a promising line of research. Furthermore, exploring the ways to constrain the output of the algorithm to the snowfall retrievals by the CloudSat radar may also help to improve the accuracy of snowfall detection. A physically realistic definition of mixed-phase precipitation based on cloud microphysics may reduce the uncertainties in phase retrievals. Finally, future research is also required to expand and evaluate the proposed algorithm with direct comparison of its results with ground-based gauge observations.

\newpage
\acknowledgments{The authors acknowledge the support from the National Aeronautics and Space Administration (NASA) through a Precipitation Measurement Mission award (NNX16AO56G) and a New Investigator Program award (80NSSC18K0742). Zeinab Takbiri acknowledges the support provided by the Minnesota's Discovery, Research, and Innovation Economy (MnDRIVE 2017) fellowship. Pierre-Emmanuel Kirstetter acknowledges support from the NASA Precipitation Science Program (NNX16AE39G) and from the GPM mission Ground Validation Program (NNX16AL23G). The contributions from F. Joseph Turk were performed at the Jet Propulsion Laboratory, California Institute of Technology, under a contract with NASA. The GPM data (version 4) is provided courtesy of the NASA Precipitation Processing System at the Goddard Space Flight Center (\url{https://pmm.nasa.gov/data-access/}). The MERRA-2 and MODIS data are from the Goddard Earth Sciences and Information Service Center (\url{https://disc.sci.gsfc.nasa.gov/mdisc/}) and the Land Processes Distributed Active Archive Center by the USGS (\url{https://lpdaac.usgs.gov/data_access/data_pool}). The authors would also like to thank Ryan Currier from the University of Washington for providing the data from the Weather Research and Forecasting Model over the Olympic Mountains.}

\bibliography{biblio}
\end{document}